\documentclass[aps,pra,amsmath,amssymb,amsthm,floatfix,nofootinbib,notitlepage,superscriptaddress,10pt]{revtex4-2}

\usepackage{esvect}
\usepackage{mathtools}
\usepackage{graphicx,xcolor}
\usepackage{todonotes}
\usepackage{siunitx}

\def\actionC{\mathcal{I}}

\def\actionF{\mathcal{I}_F}
\usepackage{float}
\pdfoutput=1
\usepackage{amsthm}
\usepackage{latexsym}
\usepackage{amsfonts}
\usepackage{bbm,dsfont}
\usepackage{graphicx,xcolor}
\usepackage{braket}
\usepackage{hyperref}
\usepackage[normalem]{ulem}
\usepackage{braket}
\definecolor{bottle_green}{RGB}{0,106,78}
\definecolor{celadon_green}{RGB}{47,132,124}
\definecolor{emerald}{RGB}{80,220,100}
\definecolor{jade}{RGB}{0,168,107}
\hypersetup{colorlinks,
linkcolor=bottle_green,
citecolor=celadon_green,
urlcolor=emerald,
anchorcolor=jade}

\newcommand{\beq}{\begin{equation}}
\newcommand{\eeq}{\end{equation}}
\newcommand{\Fu}{F}

\def\eps{\epsilon}
\begin{document}
\title{Anomalous contribution to galactic rotation curves due to stochastic spacetime }
\author{Jonathan Oppenheim}
\affiliation{Department of Physics and Astronomy, University College London, Gower Street, London WC1E 6BT, United Kingdom}
\author{Andrea Russo}
\affiliation{Department of Physics and Astronomy, University College London, Gower Street, London WC1E 6BT, United Kingdom}
\begin{abstract}
   We consider a proposed alternative to quantum gravity, in which the spacetime metric is treated as classical, even while matter fields remain quantum. Consistency of the theory necessarily requires that the metric evolve stochastically. Here, we show that this stochastic behaviour leads to a modification of general relativity at low accelerations. 
    In the low acceleration regime, the variance in the acceleration produced by the gravitational field is high in comparison to that produced by the Newtonian potential, and can act as an entropic force, causing a deviation from Einstein's theory of general relativity. We show that in this ''diffusion regime'', the entropic force acts from a gravitational point of view, as if it were a contribution to the matter distribution. 
    We compute modifications to the expectation value of the metric via the path integral formalism, and find a stochastic contribution which corresponds to a cosmological constant, anti-correlated with a contribution which has been used to fit  galactic rotation curves without  dark matter. We caution that a greater understanding of this effect is needed before conclusions can be drawn, most likely through numerical simulations, and provide a template for computing the deviation from general relativity which serves as an experimental signature of the Brownian motion of spacetime.   
\end{abstract}
\maketitle

According to the standard model of cosmology, $\Lambda$CDM,  visible matter makes up only 5\% of its contents, with dark energy or a cosmological constant $\Lambda$  and cold dark matter (CDM) making up the remaining part. There is strong evidence for this. Dark energy or a cosmological constant appears to drive the expansion of the universe, while dark matter can account for the flatness of galactic rotation curves~\cite{rubin1978extended}. It is observed in the CMB power-spectrum~\cite{ade2016planck,hinshaw2009five}, by gravitational lensing~\cite{taylor1998gravitational} such as that observed in the Bullet Sluster, through dispersion relations of elliptical galaxies~\cite{faber1976velocity}, mass estimates of galaxy clusters~\cite{allen2011cosmological},
and appears to be required for the formation of galaxies in the early universe~\cite{Conselice_2014}. However, despite large scale efforts, neither dark energy nor dark matter have been directly detected. Their apparent existence is only felt through their gravitational field.  Discoveries in physics are often indirect. The neutrino was conjectured by Pauli to exist in 1939 in order to account for energy conservation in $\beta$-decay, and only gave a signal in a particle detector 26 years later. But in the absence of any direct evidence for dark energy or dark matter it is natural to wonder whether they may be unnecessary scientific constructs
like celestial spheres, ether, or the planet Vulcan, all of which were superseded by simpler explanations. Gravity has a long history of being a trickster.

Several attempts to modify gravity without dark matter have been proposed. In 1983 Milgrom~\cite{milgrom1983modification} found that if a theory had the property that either the law of inertia were modified, or Newton's theory of gravity was modified at low acceleration such that
\begin{align}
a = 
\begin{cases} 
	a_N & \text{when } a \gg a_0, \\
	\sqrt{a_0 a_N} & \text{when } a \ll a_0.
\end{cases}
\label{eq:MONDish}
\end{align}
with $a_N$ the Newtonian acceleration, and $a_0$ a parameter of order $10^{-10}$ $m/s^2$, then  the flatness of rotation curves and the Tully-Fischer relation~\cite{TullyFisher} would follow. These are effects currently attributed to dark matter. He called this Modified Newtonian Dynamics (MOND)~\cite{sanders2002modified,bekenstein2006modified,milgrom2008mond}. Here, we would like to point out that this behaviour is reminiscent of Brownian motion, with a mean of $a_N$, and standard deviation of $\sqrt{a_N}$, a statement which we hope becomes clearer as we progress. 
In 1983 Milgrom also observed~\cite{milgrom1983modification} that the MOND acceleration (in the units of this article, $c=\hbar=1$), is given by 
   \begin{equation}
        a_0\approx\frac{1}{2\pi}\sqrt{\frac{\Lambda}{3}},
        \label{eq: coincidence}
    \end{equation}
a coincidence that has yet to be explained.  A number of theories have been proposed to reproduce MOND phenomenology~\cite{bekenstein2006modified,milgrom2008mond,jusufi2024apparent,torrome2024emergence,Finster_2024,Milgrom_1999,mondparadigm2024}, 
but thus far no satisfying fundamental theory which reproduces this behaviour has been found. The problem is that while it is easier to modify a theory at high energy, modifying a theory at low energy while still respecting current experimental bounds is difficult. It is also important to emphasise that MOND has yet to account for the results of gravitational lensing or the CMB power spectrum. However, it seems reasonable to wait regarding these, because MOND is not a fundamental theory.

Another approach, initiated by Mannheim, comes from fitting rotation curves to the spherically symmetric metric which is a solution to conformal gravity~\cite{mannheim1997galactic,mannheim1989exact,mannheim2012fitting}, an approach we will discuss in more detail after having presented the results of the path integral. 
Here, we calculate the effect on rotation curves due to a recently proposed alternative to quantum gravity~\cite{oppenheim2018post,oppenheim2023covariant}. We will, in particular, use the path integral formulation developed in~\cite{oppenheim2023path,oppenheim2023covariant}, with Zach Weller-Davies.  
The theory was not developed to explain dark matter or energy, but rather, to reconcile quantum theory with gravity. However, it was already noted in~\cite{oppenheim2018post}, that diffusion in the metric could result in stronger gravitational fields when one might otherwise expect none to be present, and that this raised the possibility that gravitational diffusion may explain galactic rotation curves and galaxy formation without the need for dark matter. Here, we will add weight to this intuition. 

We will find that even when the bare cosmological constant is zero, one should typically expect a small one due to stochastic fluctuations. This is intriguing because the bare cosmological constant is taken to be zero when no matter is present, to preserve positivity of correlation functions~\cite{grudka2023renormalisable}. This provides a potential explanation for its small but non-zero value. 
We further find that stochastic fluctuations act as if they are a positive contribution to the mass, and that these become relevant at an acceleration scale. This sets a regime at which gravity is modified, which is a necessary condition for MOND behaviour.  This is discussed through Eq.~\eqref{eq: NewtVar}. 

The path integral also contains another dominant anomalous contribution, which in the spherically symmetric case, contributes on average a linear term in the radial direction, $\gamma_1 r$ to the Schwarzschild and Newtonian solution, akin to that found in conformal gravity. $\gamma_1$ has units of acceleration, and could arise through the dynamics of the theory.
 
Although the theory does not precisely predict the coincidence of Eq.~\eqref{eq: coincidence}, it gives a numerically similar relationship between the fluctuation corresponding to the acceleration $\gamma_1$ of
$\gamma_1\approx\Lambda R_H$, with $R_H$ the Hubble radius and with the relationship becoming $\gamma_1\approx\sqrt{\Lambda}$ in a $\Lambda$-dominated universe. A correlation between a cosmological constant type term, and the acceleration term, is a prediction of the path integral, and could help explain the coincidence problem, namely that the energy density of the dark energy and dark matter contributions are roughly equal. The linear term
results in flat rotation curves in a region far from the galactic centre,  with possible deviations at larger distances depending on how parameters are chosen. Our parameters are not coupling constants, but correspond to boundary conditions which likely require a fuller understanding of the dynamics of the theory. Since quantum-classical theories of gravity are very restricted, a better understanding of this is likely to enable astrophysical tests of the quantum nature of spacetime. 

The theory of ~\cite{oppenheim2018post,oppenheim2023covariant} describes spacetime classically, while the matter fields are quantised. This necessarily requires the evolution of spacetime to have a stochastic component~\cite{oppenheim2023gravitationally}. The full path integral of the theory is given in Section \ref{app: CQ}. Here, we only study the classical limit of the theory. 
In this limit, the quantum
matter degrees of freedom will have decohered, but in the dynamics of the classical-quantum framework, the classical degrees of freedom still undergo stochastic evolution. 
We will also not concern ourselves with the evolution of the matter degrees of freedom, and thus only represent them by their mass density $m(x)$, neglecting the Hamiltonian term which governs their evolution.
Therefore, we can represent the path integral from time $t=0$ to $t_f$ as
\begin{equation}
    \label{eq: Newton_decohered_PI}
    \varrho(\Sigma_f,m_f,t_f)= \int \mathcal{D}g  \;\mathcal{N} e^{\actionC[g,m,0,t_f]}  \varrho(\Sigma_i,m_i,0), 
    \end{equation}
    where the action contains a gauge-fixing term so that the path integral over metrics $g$ is over geometries. The path
integral determines the probability density $\varrho(\Sigma_f,m,t_f)$ of a final spatial surface $\Sigma_f$ given an initial spatial surface,
and $\mathcal{N}$ is a normalisation factor. This is much simpler in the Newtonian limit, where we can parameterise the metric
in terms of the Newtonian gravitational potential $\Phi$. In this limit, the action of~\cite{oppenheim2023covariant} was found to be~\cite{layton2023weak}
\begin{equation}
\label{eq:main-isotropic_zero_order}
    \actionC[\Phi,m,0,t_f] = -\frac{D_0(1-\beta)}{G_N^2}\int_{0}^{t_f} dtd\vec{x}\, \big( \nabla^2 \Phi - 4\pi G_N m(x)\big)^2 .
\end{equation}
Here $\beta$ is required to be less than 1/3 by consideration of positivity of the full action, but for the correlation functions of the full theory to be positive semidefinite, it is required to be negative~\cite{grudka2023renormalisable}. It is naturally assumed to be of order $\mathcal{O}(1)$, and in this Newtonian limit, $\beta<1$ is clearly sufficient -- it is what is required for the path integral to suppress paths away from Poisson's equation. $D_0/G_N^2$ is a dimensionless coupling constant, which determines the scale of fluctuations. $\beta=1/3$ without the matter couplings, would correspond to the conformally invariant theory. The action is similar in form to that of the Onsager-Machlup function~\cite{Onsager1953Fluctuations}. The action can also be derived as the weak field limit of  the action for Nordstrom gravity which is diffeomorphism invariant and completely positive~\cite{UCLNordstrom}.

Crucially, we see that the action \eqref{eq:main-isotropic_zero_order} is in the form of an equation of motion squared, and has a global maximum when the equations of motion are satisfied
\begin{equation}
   \langle \nabla^2\Phi -4\pi G_N m\rangle=0.
    \label{eq:ExpectPoisson}
\end{equation}
As shown in~\cite{layton2023weak}, this action derived as the weak field limit of~\cite{oppenheim2018post,oppenheim2023covariant} is a path integral formulation of the model of ~\cite{tilloy2016sourcing} when a local noise kernel is chosen.
Since Eq \eqref{eq:ExpectPoisson} is linear in $\Phi$, when $m(x)$ has a definite distribution rather than being a statistical mixture of distributions we have that $\nabla^2\langle\Phi\rangle$ satisfies Poisson's equation, and so on expectation, there is no difference between the expectation value of $\Phi$ and its deterministic value.
Nonetheless, the action of Eq \eqref{eq:main-isotropic_zero_order} is extremised not only by $\Phi$ which satisfies Poisson's equation, but also by more general field configurations that make the action variation vanish for fixed endpoints. In vacuum, when $m(x)=0$, the $\Phi$ which extremise the action of Equation \eqref{eq:main-isotropic_zero_order} is found to be the biharmonic equation
 \begin{equation}
    \nabla^4\Phi=0,
    \label{eq:biharmonic}
\end{equation}
which have solutions away from $x=0$ given by  
\begin{align}
	\Phi_{MPP}(x)&= -\frac{\kappa_m}{4\pi|x|} + \kappa_0 - \kappa_1 8\pi |x|  + \kappa_2 |x|^2 .
\label{eq:phimpp}
\end{align}
The first two terms are the standard Newtonian potential plus an arbitrary constant term, while the last two additional terms do not satisfy the standard vacuum Poisson's equation, and are therefore local rather than global maxima. However, they still make substantial contributions to the path integral. Note that the $\kappa_m$ term and the $\kappa_1$ term are Green's functions for $\nabla^2$ and $\nabla^4$ respectively,  and for this reason we've explicitly put in the sign and factors of $\pi$. Solutions to the biharmonic equation with a source can be found for example, in \cite{mannheim1994newtonian}, the difference here being that the source
is $4\pi\nabla^2\Phi$.

We wish to emphasise that Equation \eqref{eq:phimpp} is not equations of motion in the usual sense.  The dominant contribution to the path integral comes from the solution to Poisson's equation, while the rest merely represent stochastic deviations from Poisson's equation which are not too suppressed in the path integral given boundary conditions. Other configurations also contribute with a probability weighed by the path integral.

Therefore, we will call the generalised configurations such as those of Eq. \eqref{eq:phimpp}
 \textit{Most Probable Paths} (MPPs), adopting the language used in the study of diffusive dynamics~\cite{feynman1965quantum,durr1978onsager,chao2019onsager}.  We include a simple example of how these contribute to the path integral in the case of Brownian motion with a step function potential in Appendix~\ref{sec:Brownian}. 

More generally, what the path integral tells us is that one expects various contributions to the gravitational potential other than the solution to Poisson's equation. At short distance scales, one has fluctuations in time, which are required in order for the theory to be consistent with quantum theory, and current bounds on superpositions of massive particles. These have been discussed in detail in~\cite{oppenheim2023gravitationally,layton2023weak,UCLNordstrom}. We also expect deviations from Poisson's equation, and indeed, deviations from the full relativistic theory, due to non-linearities in the full path integral. These we will discuss in relation to Eq.~\eqref{eq: NewtVar}. Finally, we should also expect deviations arising due to the theory's dynamics. These could include early time temporal fluctuations which get stretched out during the expansion of the universe as happens with vacuum fluctuations of quantum fields, or fluctuations which are built up over cosmological or galactic time scales. These will be discussed further in~\cite{UCLcosmo}. Here we will study the later two types of fluctuations. In this case, if the dynamics is slow enough, one expects the action of Eq.~\eqref{eq:main-isotropic_zero_order}  to characterise the final distribution of deviations away from Poisson's equation, but without the integral over time. Thus,
unlike the Brownian motion path integral in Appendix~\ref{sec:Brownian}, we first treat this as a non-dynamical path integral, characterising the relative probabilities of various deviations away from the Newtonian potential, regardless of how they arose. Indeed, paths such as $\Phi_{MPP}(x)$ in Eq.~\eqref{eq:phimpp} should be treated as static as they are inherited from the relativistic theory and we would not expect them to fluctuate over time. 
We, therefore, drop the integral over $dx^0$. To keep track of units, we replace the dimensionless coupling constant $D_0/G_N^2$ with $D_{0,T}/G_N^2$ having units of distance. 

This is equivalent to breaking up the path integral in terms of short time intervals of a Lagrangian $\mathcal{L}$ 
\begin{align}
\exp{\int_0^{t_f} dt \mathcal{L} }=\exp\left[{\int_{t_f-\eps}^{t_f} dt \mathcal{L} }\right]\times \cdots \exp{\left[\int_{2\eps}^{3\eps} dt \mathcal{L} \right]}\times\exp{\left[\int_{\eps}^{2\eps} dt \mathcal{L}\right] } \times\exp{\left[\int_0^{\eps} dt  \mathcal{L}\right] }
\label{eq:breakuo}
\end{align}
and considering the case where the distribution in any time interval $\exp{\left[\int_{N\eps}^{(N+1)\eps} dt \mathcal{L} \right] }$ is roughly the same as in any other time-interval. We can therefore take $\int_{N\eps}^{(N+1)\eps}dt \mathcal{L}\approx  \mathcal{L}_T$ with $\mathcal{L}_T$ is the Lagrangian smeared over a short time interval $\eps$, and consider only one of the terms in Eqn \eqref{eq:breakuo}, since they are all approximately equal. For the action of Eqn \eqref{eq:main-isotropic_zero_order}, this amounts to taking $D_{0,T}:=D_0c\eps$ and dropping the time integral since earlier and later times don't contribute to the path integral when the distribution is stationary.  In this case, $\eps$ would need to be some minimal time over which the distribution is stationary, which is typically taken to be the response time of the system. This should be recalled when adding powers of $c$ back to the prefactor of $D_{0,T}/G_N^2$ in these units\footnote{A note about the units used in this article: The dimensionless coupling constant of the theory is $D_0 c^6/G_N^2$, see Eq. \eqref{eq: actionR}. $D_0$ is related to the decoherence rate of the theory and there are different conventions as to how many powers of $c$ are absorbed into it. In the master equation approach, such as \cite{oppenheim2023gravitationally}, $D_0$ is quoted in units $\unit{m^3/kg^2.s}$ where it is the inverse of the gravitational diffusion coefficient $D_2=1/D_0$. Here and in other path integral papers, $D_0 c^3$ is quoted in $\unit{m^3/kg^2.s}$, while $D_{0,T} c^3$ has units $\unit{m^4/kg^2.s}$. We will generally set $c=1$ until comparing our results with experiment.}. To put it another way, the full path integral computes a distribution over space and time, and we will consider these separately, first determining the marginal probability distribution over space in the case of a stationary distribution. This is valid if the dynamics is slow enough, since we can expect the system to try to minimise the spatial action and temporal action separately.

In this case, the $\kappa$ in the most probably path should be static, since this non-relativistic action follows from a local relativistic one. For now, it is worth foreshadowing our final result by noting that the contribution of $\kappa_1$ to the most probable path has units  of acceleration, and the $\kappa_2$ contribution is a solution to general relativity if there is a constant matter density and has the same units as the cosmological constant.
If we were to substitute the most probable path of Equation \eqref{eq:phimpp} into the 0th order action of \eqref{eq:main-isotropic_zero_order} in a vacuum region, then the $\kappa_m$, $m(x)$ and $\kappa_0$ term don't contribute to the action if the Newtonian term is used as a Green's function for the matter distribution $m(x)$. They can therefore be set by the boundary conditions, as is done in solving Poisson's equation. The $\kappa_2$ however, does contribute to the action and is therefore suppressed. We will neglect the $\kappa_1$ term in this brief discussion, because it doesn't satisfy appropriate boundary conditions if in vacuum, but it would still contribute significantly to the path integral (see comment in Appendix \ref{sec:comment}). If we substitute the other terms of the MPP of Eq \eqref{eq:phimpp}, into the 0th order action, we get, what we will call the {\it MPP-action}
\begin{align}
    I_{MPP}=-\frac{D_{0,T} (1-\beta)}{4\pi G_N^2}\int d^3x\left(6\kappa_2   \right)^2\quad.
    \label{eq:NewtMPPaction}
\end{align}
 When we are substituting the most probable paths, Equation \eqref{eq:NewtMPPaction} is reminiscent of the on-shell action used in quantum field theory and captures the leading order terms to the path integral. If we include other terms in the action  (such as the $\kappa_1$ term), then the action will allow us to calculate the relative probabilities of the paths we substitute into the action regardless of whether they extremise the action.

If we add a source term $J(x)O(\kappa)$ to the action with an arbitrary function $O(\kappa)$ of the parameters $\kappa=\{\kappa_m,\kappa_0,\kappa_1,\kappa_2\}$, we can construct a partition function
\begin{align}
    Z_{MPP}[J]=\mathcal{N}\int \mathcal{D}\kappa\, e^{\actionC[\Phi_{MPP},m,J]},
\end{align}
with $\mathcal{N}$ the normalisation factor. This
can be used to compute correlation functions of the parameters $\kappa$. However, for this simple example, we immediately see that this is a normal distribution in $\kappa_2$, with a standard deviation which scales like $G_N/\sqrt{D_{0,T} V}$, where $V$ is the spatial volume of the region we are considering.
We will see that $\kappa_2$ is equivalent to a small cosmological constant of arbitrary sign, and we emphasise that it appears as a necessary fluctuation even though the deterministic equations of motion don't allow for it. Since it has some variance, we would be surprised if we found it to be $0$. Care should be taken with the $MPP$ action. If we toss a 1000 coins, slightly biased towards heads, the most probably single configuration is all heads. A more natural characterisation of the outcome would be in terms of the expected number of heads vs tails, which also characterises any local sample, provided it is sufficiently large.

Let us now move to the relativistic case. Here, the static solution is the appropriate metric for considering the effect of stochastic fluctuations over large distances. 
Therefore, we consider a spherically symmetric metric of the form
  \begin{equation}
        ds^2=-e^{2\phi(r)}dt^2+e^{-2\psi(r)}dr^2+e^{-2 \chi(r)}r^2\,d\Omega^2,
        \label{eq:themetric}
    \end{equation}
with $\Omega$ the 2-dimensional solid angle. 
In general relativity, we could perform a coordinate transformation to set $\chi(r)=0$. In vacuum, Einstein's equation would require $\phi(r)=\psi(r)=\frac{1}{2}\log(1-2M/r)$ and we would arrive at the usual Schwarzschild solution. Here, we consider more general metrics of the form
\begin{align}
    \phi(r)=\psi(r)=\frac{1}{2}\log{\left(1-\frac{2\Fu(r)}{r}\right)},\quad \chi(r)=0,
    \label{eq:Fansatz}
\end{align}
with $\Fu(r)$ a function which we will take to be a power series in $r$, motivated in part by the most probable path of Equation \eqref{eq:phimpp}. 
Note that $1-2\Fu/r=0$ is a horizon, and so $\Fu$ is bounded in a similar way to the example we consider in Section \ref{sec:Brownian} of Brownian motion with a wall.

The full dynamical action of \cite{oppenheim2023covariant} is reproduced as Equation~\eqref{eq: CQaction} of Appendix~\ref{app: CQ} and in Appendix~\ref{sec:Schwar} 
we include a more detailed description of the sketch we are giving here. 
The purely gravitational part of the action, is given by
\begin{equation}
    \actionC=-\frac{D_0}{G_N^2}\int d^4x\;\sqrt{-g}\big(\mathcal{R}^{\mu\nu}\mathcal{R}_{\mu\nu}-\beta\,\mathcal{R}^2 \big).
\end{equation}
The $\mathcal{R}^2$ term, could allow for Starobinski inflation~\cite{starobinsky1980new, vilenkin1985classical,liu2018inflation}, which is favoured by CMB data~\cite{akrami2020planck}. It is also what makes the pure gravity theory renormalisable\cite{grudka2023renormalisable}.

If we now substitute the ansatz of Eq \eqref{eq:Fansatz}
into this action, we find 
 the simple form
\begin{align}
\label{eq:Faction}
 \actionF(r)
  =  &-\frac{8 \pi D_{0,T}}{G_N^2} \int  dr\, \left((1-2 \beta) \Fu''(r)^2+\frac{(4 -8 \beta)}{r^2}  \Fu'(r)^2-\frac{8 \beta }{r}  \Fu'(r)  \Fu''(r)\right),
\end{align}
where the angular part has been already integrated out so that the path integral is a two-dimensional Gaussian distribution in the variables $\Fu'$ and $\Fu''$. 
Here, $\Fu'$ has dimensions of acceleration and $\Fu''$ has dimensions of the cosmological constant. 
To see this more clearly, let us consider the power series expansion of $\Fu$
\begin{align}
\frac{2\Fu(r)}{r}&=\sum_{n=-\infty}^{\infty} \gamma_n r^{n}\nonumber\\
&=\cdots 
+\frac{\gamma_m}{r}
+\gamma_0+\gamma_1 r+\gamma_2 r^2 + \cdots,
\label{eq:powers}
\end{align}
where in the second line we have written the terms relevant to the length scales we are considering. The $\gamma_m$ ends up dropping from the action, and we will henceforth set it to $2G_N M$, since at order $r^{-1}$ it is the standard Schwarzschild term which can here be determined from boundary conditions. It makes no difference for the purposes of this discussion whether we include the $\gamma_i$ corresponding to other higher or lower powers, or  functions like  $\log(r)$. The reason for this is that the series is linear in the $\gamma_i$, so that when we substitute the expansion back into the action, we find that the coefficients follow a multivariate Gaussian distribution with zero mean and non-zero correlation.  We can then perform the Gaussian integrals over all other $\gamma_i$ we are not interested in, and the action for the remaining $\gamma_0,\gamma_1,\gamma_2$ will not change. The reason we are most interested in these contributions, is that negative powers of $r$ will not contribute far away from the mass distribution of a galaxy which is our zone of interest. On the other hand, higher powers of the expansion will be heavily suppressed by the action once the radial integral is performed, as can be verified by including them. We will see that they represent fluctuations larger than our Hubble volume. For this reason we focus  only on the correlation between $\gamma_0$, $\gamma_1$ and $\gamma_2$, and inclusion of the other $\gamma_i$ would not effect our conclusions.

If we now substitute the power series of Eq. \eqref{eq:powers} into the action and integrate from $r=0$ to some $r_{max}$, we obtain
\begin{equation}
\label{eq:OurAction}
   \mathcal{I}_\gamma=-\frac{6 \pi  D_{0,T} V}{G_N^2} \left(\frac{5-18 \beta}{r_{\max}^2} \gamma_1^2+6 (1-4 \beta) \gamma_2^2+\frac{9(1-4 \beta )}{r_{max}} \gamma_1 \gamma_2 \right),
\end{equation}
where we have dropped the constant term $\gamma_0$ for ease of presentation, as it doesn't alter the conclusions nor contribute to rotation curves. 
$V=\frac{4}{3}\pi r_{max}^3$, and we could absorb it into the coupling constant $D_{0,T}/G_N^2$ 
which renormalises it and gives it units of Planck length to the 4th power $l_p^4$, but we leave it in place to keep track of units. The analysis doesn't change much if we integrated from the inner horizon to $r_{max}$. Since we are considering large-scale fluctuations which exist over all space, they are naturally suppressed by a volume element, so it's interesting to see that $\gamma_2$ is only suppressed by this amount. We also see here, or by explicit calculation, that higher powers in the expansion of Eq.~\eqref{eq:powers} would be more suppressed, which motivated us to integrate out such higher powers. They represent fluctuations of a length scale which are not felt inside our Hubble volume. This gives the path integral over $\gamma=\{\gamma_1,\gamma_2\}$
\begin{align}
    Z_\gamma=\mathcal{N}\int \mathcal{D}\gamma e^{\actionC_\gamma[\gamma,m]}.
    \label{eq: Zgamma}
\end{align}

Integrating over 4-geometries is here limited to 4-geometries which have the metric.
\begin{align}
   ds^2 =-\left(1-\frac{2MG_N}{r}-\gamma_0-\gamma_1 r-\gamma_2 r^2\right)dt^2+\left(1-\frac{2MG_N}{r}-\gamma_0-\gamma_1 r-\gamma_2 r^2\right)^{-1}dr^2+r^2d\Omega^2
   \label{eq: stochastic Schwar}
\end{align} 
This is the {\it MK metric} of~\cite{riegert1984birkhoff,mannheim1994newtonian}, which is used to fit galactic rotation curves~\cite{mannheim1994newtonian,mannheim1997galactic,mannheim1989exact,mannheim2012fitting,O_Brien_2018}.  Here we will not fit $\gamma_1$, but instead determine it from the path integral. While this metric is a solution to conformal gravity~\cite{riegert1984birkhoff}, it is not a solution to general relativity, nonetheless, it does contribute to the classical-quantum path integral, as can be seen from Eq. \eqref{eq:OurAction}.  While conformal gravity has issues with negative norm ghosts, this is not an issue here~\cite{grudka2023renormalisable} (c.f. \cite{buccio2024physical,mannheim2020ghost,holdom2016quadratic,donoghue2021quadratic, anselmi2017quantum, salvio2018new, salvio2022bicep}). Moreover, criticisms of using the MK metric in fitting rotation curves akin to those presented in~\cite{horne2016conformal,hobson2021conformal,hobson2022conformally} are also not applicable to the classical-quantum theory of ~\cite{oppenheim2018post,oppenheim2023covariant} which is not conformally invariant, but rather scale-invariant without matter~\cite{grudka2023renormalisable}.  Scale invariance is broken by the matter action. Furthermore, although in conformal gravity, the Newtonian potential now depends on the mass distribution of the source rather than just the mass~\cite{Yoon_2013}, the correct Newtonian potential is the dominant saddle of our path integral.

For now, we see that $\gamma_2$ corresponds to the cosmological constant term of Schwarzschild deSitter, while $\gamma_1$ contributes to the geodesic equation of stars far from the galactic centre. The expectation values of $\gamma_1$ and $\gamma_2$ are zero, but the two random variables are normally distributed and inversely correlated as can be seen by inspection of Eq. \eqref{eq:OurAction} and \eqref{eq: Zgamma}. The covariance matrix of the normal distribution determined from the path integral Eq. \eqref{eq:OurAction} is 
\begin{align}
    &\Sigma_{11}=\frac{2  r_{max}^2}{3 (13-36 \beta)}\frac{G_N^2}{D_{0,T} V}  \nonumber\\
    &\Sigma_{22}= \frac{(5-18\beta)}{9(13-36 \beta ) (1-4 \beta ) }\frac{G_N^2}{D_{0,T} V} \nonumber\\
    &\Sigma_{12} = -\frac{ r_{max}}{2 (13-36 \beta)}\frac{G_N^2}{D_{0,T} V} 
    \label{eq: CovarianceMatrix-main}
\end{align}
and the conditional mean of $\gamma_1$, given an observation of $\gamma_2$ is given by
\begin{equation}
\label{eq: Conditional_mean-main}
    \mu_{\gamma_1|\gamma_2,r_{max}}
    =-\frac{9}{2}\gamma_2\, r_{max}\left(\frac{1-4 \beta }{5-18 \beta}\right),
\end{equation}
here plotted in Figure~\ref{fig: expected_gamma1}. 
We can now ask, given that we observe a value of $\gamma_2\approx\Lambda/3$, where $\Lambda\approx 10^{-52}$ $m^{-2}$ is the cosmological constant, and an $r_{max}$ given by the Hubble radius $R_H\approx 10^{26}$ $m$ over which we obesrve it, what does this tell us about the value we expect to see of $\gamma_1$? Recall that we expect $\beta$ to be negative and of order $1$, but from Fig~\ref{fig: expected_gamma1} we see that we are insensitive to it's value. In this case we find a mean value of $\gamma_1$ to be of $\mu_{\gamma_1|\gamma_2,r_{max}}\approx -10^{-26}$ $m^{-1}$. Putting units of $c$ back in, we find that $\gamma_1$ is of the order of the MOND acceleration $\gamma_1 \approx 10^{-10}$ $m/s^2$.

This is the value initially used to fit galactic rotation curves for larger spiral galaxies, in the context of conformal gravity~\cite{mannheim1989exact}.
In the region where the transition between the rising $\gamma_1 r$ term is of the order of the falling $G_NM/r$ term, the rotation curve is roughly flat. $\gamma_1$ is not too large such that it runs afoul of experimental bounds on solar system evolution. However, to account for smaller dwarf galaxies,  $\gamma_1$ is adjusted as $\gamma_1(M)=\gamma_1(1+M/10^{10} M_{\odot})$ with $\gamma_1\approx 10^{-28} m^{-1}$. Here it should be said that CDM simulations also have trouble -- the core-cusp problem\cite{moore1994evidence,deBlok:2009sp,oh2015high}, so dark matter models also require novel matter properties, additional assumptions or further considerations\cite{pontzen2014cold}. To account for more recent data, $\gamma_2$, is taken to be a $\kappa \approx 10^{-50} m^{-2}$~\cite{mannheim2012fitting} to ''flatten the curve'' while giving a slight rise which is claimed to be observed~\cite{OBrien:2011vks} (see in contrast \cite{Brouwer:2021nsr}).

Without further considering the dynamics of the matter distribution, we have no reason at this point, to select the constants in the most probable path, because they correspond to boundary conditions. It is reasonable to set $\gamma_2$ to be the value of the cosmological constant given its observation over the scale of the Hubble radius, and then use this to predict $\gamma_1$. It is not inconsistent with our obtained $\gamma_1$ for $\gamma_2\approx\kappa$ over galactic distances, since $\gamma_1\approx\mu_{\gamma_1|\gamma_2,r_{max}}$ when $r_{max}\approx 100 kpc$ given that the observed rotation curves in each galaxy only extends to a radius of that order (the galactic disk of larger galaxies). 
We will see later that there may be some cause for this apparent dual role for $\gamma_2$. However, for now, the main conclusion we would like to draw, is that the order of magnitude estimates suggest the theory makes predictions broadly in line with current observations, and suggest that simulations of the theory, combined with astrophysical observations, could be used to test its anomalous behavior. 

Efforts to understand the effect of these stochastic fluctuations in cosmology are initiated in~\cite{UCLcosmo} with Emanuele Panella and Andrew Pontzen, where we find some evidence that early time stochastic fluctuations in the gravitational degree of freedom do behave as a positive matter contribution, in terms of how they scale with the expansion factor in a Friedman-Robertson-Walker spacetime. 
Cosmology studies using different models of stochastic fluctuations have been considered in \cite{launay2024stochastic,Colas:2023wxa}. Other approaches have also tried to connect cosmology with the emergence of dark matter~\cite{kaplan2023classical} and even attempt to explain dark energy as a fluctuation of the Newtonian gravitational constant~\cite{deCesare:2016dnp}.

In order to provide a template for further comparison between models and observation, we can already get an estimate of the parameter $D_{0,T}/G_N^2$, at least at large distances if we take the cosmological constant to be the result of stochastic fluctuations. 
Since we expect that we live in a typical universe, this tells us that the variance in $\gamma_2$ should be of the order of $\Lambda^2$, so that the value of $\gamma_2$ we witness is typical. From the covariance matrix above, we can see that this sets  $D_{0,T}/G_N^2$ to be of the order of $D_{0,T}/G_N^2\approx 1/\Lambda^2V_H$, with $V_H$ the Hubble volume. 
In units with $c$ restored, it is perhaps easiest to think in terms of a diffusion coefficient 4-density 
$G_N^2/D_{0,T}c^6 V_H\approx 10^{-104}m^{-4}$.  

A fuller analysis of the variances can be found in Appendix Section \ref{app:stats}. 

Note that this could explain both the small but non-zero value of the cosmological constant, at least in terms of $D_{0,T}$, since we find in~\cite{grudka2023renormalisable} that a bare cosmological constant can't be included for reasons of complete positivity, and may lead to explanations of the coincidence that  $a_0\approx\frac{1}{2\pi}\sqrt{\frac{\Lambda}{3}}$~\cite{milgrom1983modification} -- both coincidences which has thus far received no satisfying explanation. It also may explains why general relativity is modified at low acceleration. Crucially, the fact that $\gamma_1$ and $\gamma_2$ are anti-correlated comes out of the path integral. We have also not needed to fine-tune $\beta$.

Let us now highlight the weaknesses of the calculation. To make it tractable analytically, we have restricted ourselves to spherically symmetric and static spacetimes, with metrics of the form of Eqs.~\eqref{eq: stochastic Schwar}. Allowing $\psi$ and $\phi$ to be different would be desirable. This would double the number of parameters in the action, and may give further insight into galactic rotation curves.    
A greater understanding of more general metric fluctuations does seem in order.
For simplicity we have restricted ourselves to understanding the correlation in $\gamma_1$ and $\gamma_2$. They reflect different length scales of stochastic fluctuations, but there are correlations between them, and for example, the higher powers in the expansion. 
Here, the full normal distribution reflected in Eq.~\eqref{eq:Faction} may provide some insight into the distribution of what is currently taken to be dark matter, but one must be careful since anything can be fit to a power series, and a fuller understanding of the probability distribution is required, as we don't know what other terms may contribute.
Here, the cosmological constant term serves as a reasonable candle to measure against. A fuller principle component analysis may be useful, via the Kosambi–Karhunen–Loève theorem~\cite{Love1978}. 

For spherically symmetric matter distributions, it is natural for the expectation value of the metric and its variance to be spherically symmetric. However, any realisation of the stochastic noise is highly non-uniform and is not constant in time, while the metric ansatz is constant in time and uniform over the sphere at each radius $r$. Here, we emphasise that while we would not expect temporal fluctuations to be spherically symmetric, it is natural that the expectation value of the fluctuations will be, which is what we are evaluating here.  Static spacetimes were chosen because, in the relativistic theory, we would not expect large scale fluctuations except those which are already present, and conjecture that given the scale, the anomalous contributions represent early fluctuations which have been baked in during inflation.  Evidence supporting this view is given in \cite{UCLcosmo}. The dynamics of other contributions to the path integral are unknown.

We have also only considered correlations in larger scale anomolous contributions rather than 
short distance fluctuations. We would like to better understand how these arise from the local time-dependent fluctuations present in \eqref{eq:main-isotropic_zero_order}.
We would therefor like to cross-check the results given here, with what the theory predicts for local time-dependent fluctuations. For this, it is worthwhile to look at the post-Newtonian expansion of the full theory. Here, we can see that local stochastic fluctuations lead to an acceleration scale, below which the laws of gravity are modified. This is most easily done in isotropic coordinates, which we do in Appendix~\ref{app: isotropic}. In these coordinates, the action is given by Eq.~\eqref{eq: isotropic action}
\begin{equation}
\actionC=-\frac{D_0 c^5}{64\pi^2G_N^2}\int d^3xdt\; e^{\frac{2\Phi}{c^2}}\left\{\left(\nabla^2\Phi-\frac{(\nabla\Phi)^2}{2c^2}-4e^{-\frac{2\Phi}{c^2}}\pi Gm\right)^2+\frac{3}{c^4}(\nabla\Phi)^4-4\beta\left(\nabla^2\Phi-\frac{(\nabla\Phi)^2}{2c^2}-4e^{-\frac{2\Phi}{c^2}}\pi Gm \right)^2\right\}
\label{eq: isotropic action main}
\end{equation}
where we have put powers of $c$ back in to highlight terms which contribute only at higher order.
In each term, the acceleration squared $(\nabla\Phi)^2/c^2$ plays an important role. Let us take $\beta=0$ for simplicity (the argument doesn't change much if it's non-zero), and let's also drop the $\frac{3}{c^4}(\nabla\Phi)^4$ for further simplicity -- its inclusion will only enhance the argument we are about to make. In this case, this action says that on expectation, the scalar gravitational potential $\Phi$ must satisfy
\begin{align}
	\left\langle e^{\frac{\Phi}{c^2}}\left(\nabla^2\Phi-\frac{1}{2c^2}(\nabla\Phi)^2-
 4e^{-\frac{2\Phi}{c^2}}\pi Gm(x)\right)\right\rangle=0
	\label{eq: NewtVar}
\end{align}

Here, we immediately see, that when $\langle (\nabla\Phi)^2\rangle>> \langle\nabla\Phi\rangle^2$, 
we will see on average, a deviation from the Newtonian limit of general relativity. Indeed from
Eq.~\eqref{eq: NewtVar}, we see that the extra variance acts like a positive mass term.
We call the regime when  $\langle (\nabla\Phi)^2\rangle>> \langle\nabla\Phi\rangle^2$, the {\it diffusion regime}, since when the acceleration $|\nabla\Phi|$ is small in comparison to its standard deviation, we will see a deviation from the Newtonian law of gravity. In Appendix \ref{sec:Brownian}, we define an entropic force to be just such a deviation from the deterministic equations. 
This is distinct from the entropic force used by Verlinde in the context of Holography, in which gravity itself is proposed as an entropic force which also acts as dark matter\cite{Verlinde_2017}.

If the diffusion in the acceleration is relatively constant far from the galactic center, then this naturally picks out a universal acceleration scale as occurs in MOND phenomenology. Once the acceleration drops below the level set by the diffusion in $|\nabla\Phi|$, we have a deviation from the Newtonian law, and indeed, the post-Newtonian corrections. If $|\nabla\Phi|$ is instead above the diffusion regime, the expectation value $\langle\nabla\Phi\rangle$ obeys the post-Newtonian equations of motion, which explains why PPN tests of general relativity are unaffected by the stochastic fluctuations of~\cite{oppenheim2018post,oppenheim2023covariant}.

We would like to compare the implications of our results with terrestrial tabletop experiments. Care needs to be taken, since we are trying to relate the theory at short distances to cosmological ones.  This fails for quantum field theory, which predicts a cosmological constant many orders of magnitude in comparison to what is observed. 
In order to properly relate such different length scales, we would need to better understand a number of issues which at present, remain unresolved. 
The first, is the running of the coupling constants, which can tell us how to compare the coupling constants at different length scales, and secondly, whether the fluctuations which are of interest here, have mass. If the fluctuations have an effective mass, then this suppresses fluctuations except at very short distances which could make their detection in tabletop experiments difficult. If the coupling constant $G_N^2/D_0$ runs in such a way that it becomes larger at higher energy then this points to diffusion being more relevant during inflation in comparison to the lower energies probed by table-top experiments. Preliminary research to understand these factors was initiated in~\cite{grudka2023renormalisable},
where we found that the pure gravity theory is related to that of scale invariant quadratic gravity which is asymptotically free. As is the case with QCD, in the asymptotically free case, a new length scale is set by dimensional transmutation. In quadratic gravity the running of the couplings can depend on their initial values and is computed in \cite{julve19781978quantum,fradkin1981renormalizable,benedetti2009asymtotic,buccio2024physical}. Naive argument suggests that it is the inverse of the diffusion coupling $D_0/G_N^2$ which would run to zero at short distances, and that the fluctuations are massless in the absence of matter, but whether this holds requires a more detailed calculation. An effective mass can be induced by matter couplings, or via non-linearities in the theory.
Another consideration is the IR behavior of the theory. Two different propagators have been considered for the model presented here, one in~\cite{oppenheim2023gravitationally} and one in~\cite{grudka2023renormalisable}. These are related by dimensional regularisation~\cite{morris2024dim}, which removes the IR divergence. The implications of this should be understood better. There are also factors such as the cut conditions imposed on the propagators, which can dramatically change how detectable stochastic fluctuations are at low frequency.
 
This makes relating results on cosmological scales to those at laboratory scales, contingent on these factors but we shall proceed with the assumption that there will not be a significant change in the short distance vs galactic scale behavior. Nonetheless, we should keep the caveats in mind. 

We have discussed both static anomalous contributions to the path integral, such as those given by Eqn. \eqref{eq:Faction} and temporal fluctuations which result in corrections to the path integral via Eqn. \eqref{eq: NewtVar}. The magnitude of fluctuations in the former case is controlled by $D_{T,0}$, while the latter is controlled by $D_0$. Experiments which are sensitive to temporal fluctuations, place a bound on $D_0$. Let us consider these first.

In order to relate $G_N^2/D_0$ to tabletop experiments, we should consider
the two-point correlation function of the stochastic fluctuations of $\Phi$.  From this one can compute the variance of the local acceleration. This was done in \cite{oppenheim2023gravitationally} for several two-point functions corresponding to different realisations of the theory.
Let us here consider the two-point function for this theory given in \cite{grudka2023renormalisable}. It is related to that found in the context of quadratic gravity~\cite{stelle1978classical} albeit with a different interpretation. We shall here consider the bare two-point function, without loop corrections. In the weak field limit, it is given,  by~\cite{grudka2023renormalisable}
\begin{align}
	G(x,x'):=& \langle \Phi(x)\Phi(x')\rangle-\langle \Phi(x)\rangle\langle\Phi(x')\rangle
	\nonumber\\
	=&-\frac{G_N^2}{8 \pi D_0(1-\beta)}|\vec{x}-\vec{x}'|\delta(t,t').
	\label{eq:freedom--main}
\end{align}
where $x$ is the spacetime coordinate $\vec{x},t$.
The acceleration covariance matrix can be obtained from this
\begin{align}
	\frac{\partial^2 G(x,x')}{ \partial{x_i}\partial{x'_j}}=\frac{G_N^2}{8\pi D_{0}} \left( \frac{\delta_{ij} |x - x'|^2 - (x_i - x'_i)(x_j - x'_j)}{|x - x'|^3} \right)\delta(t,t')
\end{align}
and by setting $i=j$ and summing, we obtain the variance of the acceleration $\sigma^2_a(x,x'):=\langle\nabla\Phi(x)\cdot\nabla\Phi(x')\rangle-\langle\nabla\Phi(x)\rangle\cdot\langle\nabla\Phi(x')\rangle$
\begin{align}
	\sigma^2_a(x,x')=\frac{G_N^2}{4\pi D_{0}(1-\beta)}\frac{\delta(t,t')}{|\vec{x}-\vec{x'}|}
	\label{eq:vara}
\end{align}
This is valid for low frequency modes. The full spectral density computed from the relativistic theory is given in \cite{grudka2023renormalisable} (see also Appendix \ref{sec:comparison}), and this enables one to compute the variance in acceleration per frequency. We find that the high frequency modes don't contribute since they average out, and Eq. \eqref{eq:vara} is a reasonable approximation at low frequency, for static test masses.

As discussed in \cite{oppenheim2023gravitationally}, we can use Eqn \eqref{eq:vara} to compute the variance in acceleration $\bar{\sigma}_a^2$ experienced by a test mass over some time period $2T$, by averaging the two-point function over the test mass distribution, and over the time period of the measurement. This is done in Appendix \ref{sec:comparison} where it is  related to tabletop experiments to place the above bounds on $\frac{G_N^2}{D_{0}}$. For a sphere of radius $R$ we find
\begin{align}
	\bar{\sigma}_a^2= \frac{3G_N^2}{40\pi D_{0}(1-\beta) RT}
	\label{eq:ShereSpec-main}
\end{align}
As one expects, the variance in the acceleration is larger if we look at shorter times, or in smaller regions, since the acceleration is less averaged.  
A discussion on relating the spectral density to various table-top experiments can be found in Appendix \ref{sec:comparison}.

Let us first relate bounds on $D_{0,T}$ coming from astronomical data, to bounds on $D_0$ coming from tabletop measurements of gravity. Since this is an attempt to relate stationary anomalous contributions to the Newtonian potential such as those generated by the dynamics, to local fluctuations in time, and $D_{T,0}:=\eps D_0$, we can extract the response time $\eps$.
Here, we have found that anomalous contributions to the acceleration with a standard deviation of the MOND acceleration, might explain both the small value of the cosmological constant and perhaps the flatness of galactic rotation curves. We have estimated the value of the diffusion 4-density $G_N^2/D_{0,T} V_H c^3$ which corresponds to this, to be of order $\Lambda^2$. 
This rules out terrestrial experiments being able to detect the long-range contributions discussed here, although it says little about shorter range fluctuations. Bounds on short range temporal fluctuations were calculated in \cite{oppenheim2023gravitationally} for various noise-kernels. 
In Appendix \ref{sec:comparison} we examine several experiments and show that
modern precision Cavendish experiments can put a bound on the dimensionless coupling constant given by $G_N^2/D_0c^6\leq 10^{-42}$ (or $D_0c^3\geq 10^{-2} \unit{m^3/kg^2.s}$) with some assumptions, and putting back factors of $c$. Next, we use the fact that $D_{0,T}:=D_0 c\,\eps$ for some minimum time length $\eps$, to find $\eps\leq 10^{-25}\unit{s}$. This is the characteristic timescale over which we expect the anomalous static contributions to remain static. This can then be checked against models which produce such contributions dynamically, such as those found in \cite{UCLcosmo}.
An alternative comparison undertaken in \cite{hertzberg2024comment} is critiqued in Appendix \ref{sec:comment}.

Let us now turn to Eq.~\eqref{eq: NewtVar} and attempt to relate the modification which it predicts at low acceleration, to tabletop experiments. If we regard the
$\langle(\nabla\Phi)^2\rangle$ as acting like a cosmological constant term, we would then require the variance in the acceleration $\sigma_a^2$ to be of order $\Lambda$.
This could then favour anomalous contributions in the path integral such as the $\gamma_1 r$ term due to the anti-correlation in the two variables. From Eq. \eqref{eq:ShereSpec-main}, we see the scale dependence of the modification to Newtonian gravity. What scale is relevant is not clear, since we are considering how the gravitational field reacts to its own fluctuations elsewhere. The most extreme proposition would be to consider the variance $\bar{\sigma}_a^2$ given by Eq. \eqref{eq:ShereSpec-main} when $R=H^{-1}$ is taken to be the Hubble radius and $T=H^{-1}$ is taken to be the Hubble time. This would give the smallest possible value for $\bar{\sigma}_a^2$, since we average the fluctuations over the largest possible spacetime volume, and thus require the largest possible value of $G_N^2/D_0$ for it to play a significant role as a cosmological constant. Since we are close to living in a $\Lambda$-dominated universe, we have $H^2=\Lambda/3$ and so if we take Eq \eqref{eq:ShereSpec-main} seriously over such length scales, we would require the dimensionless coupling constant $G_N^2/D_{0}$ to be of order unity. While this is certainly intriguing from a naturalness point of view, it would appear to be $40$ orders of magnitude  too large for table-top experiments. How seriously to take the extreme proposition is unclear, but its reasonable to regard it as evidence suggesting that local stochastic fluctuations are too small to contribute to the cosmological constant via their variance. Cosmological evolution occurs at such long time scales that the variances are averaged out. Whether we should also be averaging the fluctuations over the entire volume of the universe is perhaps less easy to justify, and likely requires a dynamical model to understand better, along the lines of \cite{UCLcosmo}. This anomalous contribution is very much ''observer dependent'' in the sense that it depends on the time and length scale over which other systems interact with the fluctuations. On shorter time scales and distances, one has a diffusion regime which is close to the MOND scale, and we discuss this in relation to current experimental bounds in Appendix \ref{sec:comparison}.

We can also use the decoherence vs diffusion trade-off~\cite{oppenheim2023gravitationally} to relate bounds on gravitational diffusion to terrestrial decoherence experiments. The trade-off, is a requirement on any classical-quantum theory, and says that the amount of diffusion in the gravitational field can be bounded by the decoherence rate of superpositions of massive objects.
In \cite{oppenheim2023gravitationally}, it was found that the decoherence rate corresponding to the path integral of Eq.~\eqref{eq: CQaction} or its weak field limit~\cite{layton2023weak} to be $\lambda=2D_{0}M^2/V_\lambda$, where $M$ is the mass of the particle in the interference experiment, and $V_\lambda$ is the volume of the wave-packet. Note that this decoherence rate is not the Diosi-Penrose rate~\cite{karolyhazy1966gravitation,diosi1989models,penrose1996gravity}, since the theory considered here is ultra-local and linear. There is also no genuine decoherence, since the quantum state stays pure, conditioned on the classical degrees of freedom~\cite{layton2023weak}, here taken to be spacetime. Only when integrating out the gravitational degrees of freedom does the state decohere in this theory.  
Using $M\approx 10^{-24}$ kg for fullerene molecules, $V_\lambda\approx 10^{-25}$ $m^3$ for the wavepacket volume estimated in the experiment of~\cite{Gerlich2011} and a decoherence rate $\lambda\geq .1 \unit{s^{-1}}$~\cite{Gerlich2011},
gives $G_N^2/D_0c^6 \geq 10^{-60}$($D_0c^3\leq 10^{24} \unit{m^3/kg^2 s}$).
There is also secondary decoherence due to the fluctuations of spacetime\cite{tilloy2017principle}, since temporal fluctuations contribute a random contribution to the Hamiltonian when the gravitational field is integrated out. These are included in the discussion in Appendix \ref{sec:comparison}. Combining decoherence experiments and precision tests of gravity gives a decoherence vs diffusion squeeze of
$10^{-2}\leq D_0c^3\leq 10^{24} \unit{m^3/kg^2 s}$. This theory is thus currently viable, with the proviso that the IR regularisation needs to be better understood. We estimate that the lower bound can be increased by at least $10$ orders of magnitude once LISA launches, and expect that the upper bound can be lowered to match it. Other 
experiments to test the quantum vs classical nature of spacetime are also becoming feasible~\cite{bose2017spin,marletto2017gravitationally,oppenheim2023gravitationally, lami2023testing,kryhin2023distinguishable,carney2019tabletop}.

At this point, it is too early to make bold claims in either direction, and a greater understanding of the theoretical and experimental constraints is required. It is also possible that the effects derived here could be the result of a fully quantum theory of gravity, which~\cite{oppenheim2018post,oppenheim2023covariant} describes as an effective theory. This we regard as unlikely because we don't expect the stochastic fluctuations of spacetime to be so large in a quantum theory of gravity. The parameter space of such an effective theory has been found with Isaac Layton in~\cite{layton2023classical}. We also do not know how to make such a theory renormalisable. Baring such a theory, it would appear that 95\% of the energy in the universe is due to stochastic fluctuations of spacetime, whose origin is either due to a fundamental breakdown in predictability or an environment which does not obey the laws of classical or quantum theory~\cite{oppenheim2023soup}.

While this study demonstrates that galactic rotation curves can undergo modification due to stochastic fluctuations, a phenomenon attributed to dark matter, it is important to acknowledge the existence of separate, independent evidence supporting $\Lambda$CDM. In particular, in the CMB power spectrum, in gravitational lensing, in the necessity of dark matter for structure formation, and in a varied collection of other methods used to estimate the mass in galaxies. These now form an important set of tools with which to test the theory of \cite{oppenheim2018post,oppenheim2023covariant}.

{\bf \noindent Acknowledgements:} We are grateful to  Maite Arcos, Isaac Layton, Emanuele Panela, Andrew Pontzen, Zach Weller-Davies, Keith Horne, Stacy McGaugh, Philip D. Mannheim, Mordehai Milgrom, Geoff Pennington, Herman Verlinde, Erik Verlinde, C.D. Hoyle, Dan Carney, and Gerard Milburn for valuable discussions. JO was supported by an EPSRC Established Career Fellowship, and by the It from Qubit Network grant from the Simons Foundation. AR was supported by a grant from UKRI.

\bibliography{refcq,NewtLimitbib,refRenorm}

\newpage
\appendix
\section{The classical-quantum action in the classical limit.}
\label{app: CQ}
Examples of consistent ways to couple classical and quantum systems via a master-equation approach have been known since the 90's~\cite{blanchard1993interaction,blanchard1995event,diosi1995quantum}. One can derive the most general form of consistent classical-quantum (CQ) dynamics, by
demanding that the dynamics preserves the split of classical and quantum degrees of freedom, and preserve the positivity and normalisation of probabilities~\cite{oppenheim2018post,UCLPawula}. This can then be used to construct a master equation for general relativity via the Hamiltonian formulation~\cite{oppenheim2018post,UCL2022constraints,oppenheim2021constraints}. Recently a path integral formulation of classical-quantum dynamics was introduced with Zach Weller-Davies~\cite{oppenheim2023path} and used to formulate a manifestly covariant path integral for classical general relativity coupled to quantum fields~\cite{oppenheim2023covariant}.
A measurement and feedback approach in the case of sourcing the Newtonian potential by quantum matter has also been pursued~\cite{kafri2014classical,tilloy2016sourcing,tilloy2017principle}, as well as an unravelling approach~\cite{UCLunrav,UCL2022semi}.
These can be applied to the weak field limits of gravitational theories~\cite{layton2023weak,UCLNordstrom} and to cosmology~\cite{UCLcosmo}.

In the present article, we don't need the full CQ path integral, since we are interested in the limit where the matter fields behave classically. But for completeness, 
we present the full path integral of \cite{oppenheim2023covariant}
\begin{equation}
	\label{eq: GRPathMain}
	\varrho(g_f,\phi^+_f,\phi^-_f,t_f)= \int \mathcal{N} \mathcal{D} g\mathcal{D}\phi^+\mathcal{D}\phi^-  \; e^{\mathcal{I}_{CQ}[g,\phi^+,\phi^-_,t_i,t_f]}  \varrho(g_i,\phi^+_i,\phi^-_i,t_i) ,
\end{equation}
where $\mathcal{N}$ is a normalisation factor and the action takes the form of:
\begin{equation}
	\label{eq: CQaction}
	\begin{split}
		\mathcal{I}_{CQ}[g,\phi^+,\phi^-,t_i,t_f] &=\int_{t_i}^{t_f} d^4x\, \bigg[ i\big(\mathcal{L}_{Q}[g,\phi^+]-\mathcal{L}_{Q}[g,\phi^-]\big)\\&\quad -\frac{\text{Det}[-g]}{8}\big(T^{\mu\nu}[\phi^+] - T^{\mu \nu}[\phi^-]\big) D_{0,\mu \nu\rho \sigma}[g]\big(T^{\rho\sigma}[\phi^+] - T^{\rho \sigma}[\phi^-]\big) \\
		& \quad - \frac{\text{Det}[-g] c^6}{128\pi^2 G_N^2 } \left( G^{\mu \nu} - \frac{8 \pi G_N}{c^4}\bar{T}^{\mu\nu}[\phi^+,\phi^-]  \right) D_{0, \mu \nu \rho \sigma}[g] \left( G^{\rho \sigma} - \frac{8 \pi G_N}{c^4}\bar{T}^{\rho\sigma}[\phi^+,\phi^-]  \right)
		\bigg].
	\end{split}
\end{equation}
Here $\mathcal{L}_{Q}$ is the quantum Lagrangian density including the appropriate metric factors, the bra and ket fields $\phi^\pm$ can be any quantum fields, $\bar{T}[\phi^+,\phi^-]$ is the average of the bra and ket fields of the stress-energy tensor and we have taken $D_0$ to saturate the decoherence-diffusion trade-off \cite{oppenheim2023gravitationally} such that both the decoherence and diffusion coefficients are written in terms of $D_0$.  
The bare cosmological constant must be taken to be zero for the action to be completely positive~\cite{grudka2023renormalisable}.  Since we do not consider the dynamics of the matter distribution, and consider the decohered case when $T^{\mu\nu}[\phi^+]=T^{\mu\nu}[\phi^-]$, only the final line of the action is used here, and $\bar{T}^{\rho\sigma}[\phi^+,\phi^-]$ can be replaced by the classical stress-energy tensor.

The decoherence and diffusion coefficient $D_{0,\mu\nu\rho\sigma}[g]$ is then chosen to be ultra-local so that it can be written in terms of the generalised deWitt metric~\cite{dewitt_1967,giulini_kiefer_1994}:
\begin{equation}
	\label{eq: DeWitt_decoherence}
	\begin{split}
		&D_{0,\mu \nu \rho \sigma}=64 \pi^2\frac{D_0}{\sqrt{-g}}\big(\,g_{\mu\rho} g_{\nu \sigma}+g_{\mu \sigma} g_{\nu \rho}-2\beta g_{\mu\nu} g_{\rho \sigma}\big).
	\end{split}    
\end{equation}
where we have renormalised the constant to absorb the factor of $128 \pi^2$ in order to simplify the calculations. 
It was found in \cite{grudka2023renormalisable} that positivity requirements impose $\beta \leq\frac{1}{3}$ and, if one requires the propagator to be positive, one will have to take $\beta < 0$. In the absence of mass, the value of $\beta = \frac{1}{3}$ would correspond to conformal gravity.

In the absence of matter and energy, and with the choice of diffusion coefficient delineated in Eq.~\eqref{eq: DeWitt_decoherence}, the action of Equation~\eqref{eq: CQaction} is reduced to a purely diffusive term in the metric degrees of freedom. The first two lines of the action of Eq.~\eqref{eq: CQaction} don't contribute, and with the stress-energy being zero, we are left with
\begin{equation}
	\mathcal{I}[g]=-\frac{ D_0 c^6}{G_N^2}\int d^4x\, \sqrt{-g} \Big(G^{\mu\nu}G_{\mu\nu} - \beta\, G^2 \Big),
	\label{eq: actionG}
\end{equation}
where $G$ is the trace of the Einstein tensor. Next, we used the identities connecting the Einstein and the Ricci tensor in dimension $4$.
\begin{align}
	&G^{\mu\nu}G_{\mu\nu}=\mathcal{R}^{\mu\nu}\mathcal{R}_{\mu\nu}, \nonumber\\
	&G^2=\mathcal{R}^2,
\end{align}
to write the action as
\begin{equation}
	\begin{split}
		\mathcal{I}[g]=-\frac{ D_0 c^6}{G_N^2}\int d^4x\;\sqrt{-g}\Big(\mathcal{R}^{\mu\nu}\mathcal{R}_{\mu\nu}-\beta\,\mathcal{R}^2 \Big).
		\label{eq: actionR}
	\end{split}
\end{equation}
This action is related to that of quadratic gravity~\cite{stelle1977renormalization,stelle1978classical,SalvioQuadratic18,donoghue2021quadratic}, and is thus renormalisable, and doesn't suffer from negative norm ghosts~\cite{grudka2023renormalisable}. As far as we know, it is unrelated to other forms of stochastic gravity~\cite{hu2008stochastic} (c.f. \cite{moffat1997stochastic}) whose purpose is to approximate the quantum stress-energy tensor beyond the semi-classical regime (c.f. \cite{UCL2022semi}).

One caveat we wish to highlight, concerns the negative definiteness of this action. This is required in order to give finite probability distributions, and suppress paths which deviate from Einstein's equation. In the present article, we concern ourselves with the weak field limit of the theory, which has this property. However, the generalised deWitt metric itself, Eq.~\eqref{eq: DeWitt_decoherence}, is not positive semidefinite, nonetheless the negative contributions to the path integral appear to correspond to non-dynamical degrees of freedom~\cite{grudka2023renormalisable,UCLnorm2023}. One corresponds to the Gauss-Bonnet term, which in $4$ spatial dimensions is a purely topological term and also a total divergence. Since we don't sum over topologies, its bulk contribution is benign. The total divergence is usually discarded as a boundary term at spatial and temporal infinity which does not effect local physics, but whether this can be done here is less clear, since the final condition is not determined by the initial condition. The other negative contribution, corresponds to the magnetic part of the Weyl curvature which is also non-dynamical, in the sense of being made up of only first time-derivatives in the metric. In~\cite{UCLnorm2023}, we find there are discretizations of the path integral, such that the magnetic Weyl term merely contributes to the normalisation, and thus appears benign, but the consequences of this are not yet fully understood. This is briefly previewed in~\cite{grudka2023renormalisable}. This concern doesn't effect the calculation here, because the Weyl curvature term is positive definite on the metrics we consider, and comes into the action with an overall minus sign if the Gauss-Bonnet identity is used. Nonetheless, care should be taken in extending this work to dynamical spacetimes~\cite{barrow2002weyl} until this issue is better understood.

\section{A stochastic modification to the Schwarzschild metric}
\label{sec:Schwar}

We start with the most general spherically symmetric metric
\begin{equation}
	ds^2=-e^{2\phi}dt^2+e^{-2\psi}dr^2+e^{-2 \chi}r^2(d\theta^2+sin^2(\theta)d\phi^2).
\end{equation}
When deriving Schwarzschild, one usually redefines $r$ to reduce the metric to two free parameters before using Einstein's equation. We will do that here for simplicity, but it's important to note that this sort of coordinate system is not completely sensible here, since the metric is undergoing stochastic changes which would require one to constantly redefine $r$ to obtain a strictly static object. However, we should be able to redefine $r$ to remove $e^{2\chi}$ on expectation.

Let us consider metrics of the generalised Schwarzschild form 
\begin{align}
	\phi(r)=\psi(r)=\frac{1}{2}\log{\left(1-\frac{2\Fu(r)}{r}\right)}.
	\label{eq: ansatz-app}
\end{align}
Given this choice of metric, the curvature terms appearing in the diffusion action are:
\begin{align}
	&G^{\mu\nu}G_{\mu\nu}=\mathcal{R}^{\mu\nu}\mathcal{R}_{\mu\nu} = \frac{2}{r^2}F''(r)^2+\frac{8}{r^4}F'(r)^2,\\
	&G^2=\mathcal{R}^2 = \frac{4}{r^2}\left( \nabla^2F(r)\right)^2,
\end{align}
which when inserted in Eq.~\eqref{eq: actionG} leads to the action
\begin{align}
	\mathcal{I}=-\frac{16 \pi D_{0,T}}{G_N^2} \int dr  \left(\frac{1}{2}\Fu(r)''^2+\frac{2}{r^2}\Fu(r)'^2-\beta(\nabla^2\Fu(r))^2 \right),
\end{align}
where we have already integrated over the angular part given that the action is spherically symmetric.

We now consider the power expansion of $\Fu(r)$ of Eq. \eqref{eq:powers}  and substitute it back into the action to obtain 
\begin{equation}
	\mathcal{I}_\gamma= -\frac{ 8\pi  D_{0,T}}{G_N^2}\int_0^{r_{max}} dr \left((5-18\beta) \gamma_1^2+18 (1-4\beta) \gamma_2^2 r^2+18 (1-4\beta) \gamma_1 \gamma_2 r\right),
\end{equation}
which, when integrated, gives
\begin{equation}
	\mathcal{I}_\gamma=-\frac{6 \pi   D_{0,T} V}{G_N^2} \left((5-18 \beta) \frac{\gamma_1^2}{r_{max}^2}+6 (1-4 \beta) \gamma_2^2+9 (1-4 \beta ) \frac{\gamma_1 \gamma_2}{r_{max}}\right).
\end{equation}
where $V=\frac{4}{3}\pi r_{max}^3$. We can now see that the exponentiated action now represents a bivariate normal distribution for the parameters $\gamma_1$ and $\gamma_2$. This represents the values of the stochastic contributions to the path integral.
\begin{equation}
	\Phi =  -\frac{G_N M}{r}-\frac{\gamma_0}{2}-\frac{\gamma_1}{2} r-\frac{\gamma_2}{2} r^2.
\end{equation}
When inserted into the path integral of Eq. \eqref{eq: Zgamma}, it computes the normalised probability distribution to:
\begin{equation}
	\label{eq: PDF}
	\begin{split}
		&f(\gamma)=\frac{1}{\mathcal{N}}\,\exp\bigg(-\frac{6 \pi   D_{0,T} V}{G_N^2} \left((5-18 \beta) \frac{\gamma_1^2}{r_{max}^2}+6 (1-4 \beta) \gamma_2^2+9 (1-4 \beta ) \frac{\gamma_1 \gamma_2}{r_{max}}\right)\bigg), \\
		&\mathcal{N}=\frac{ r_{max}}{3\sqrt{3(1-4\beta)(13-36\beta)}}\frac{G_N^2}{ D_{0,T} V}.   
	\end{split}
\end{equation}
which can be seen in the contour plot:
\begin{figure}[H]
	\centering
	\includegraphics[width=0.8\textwidth]{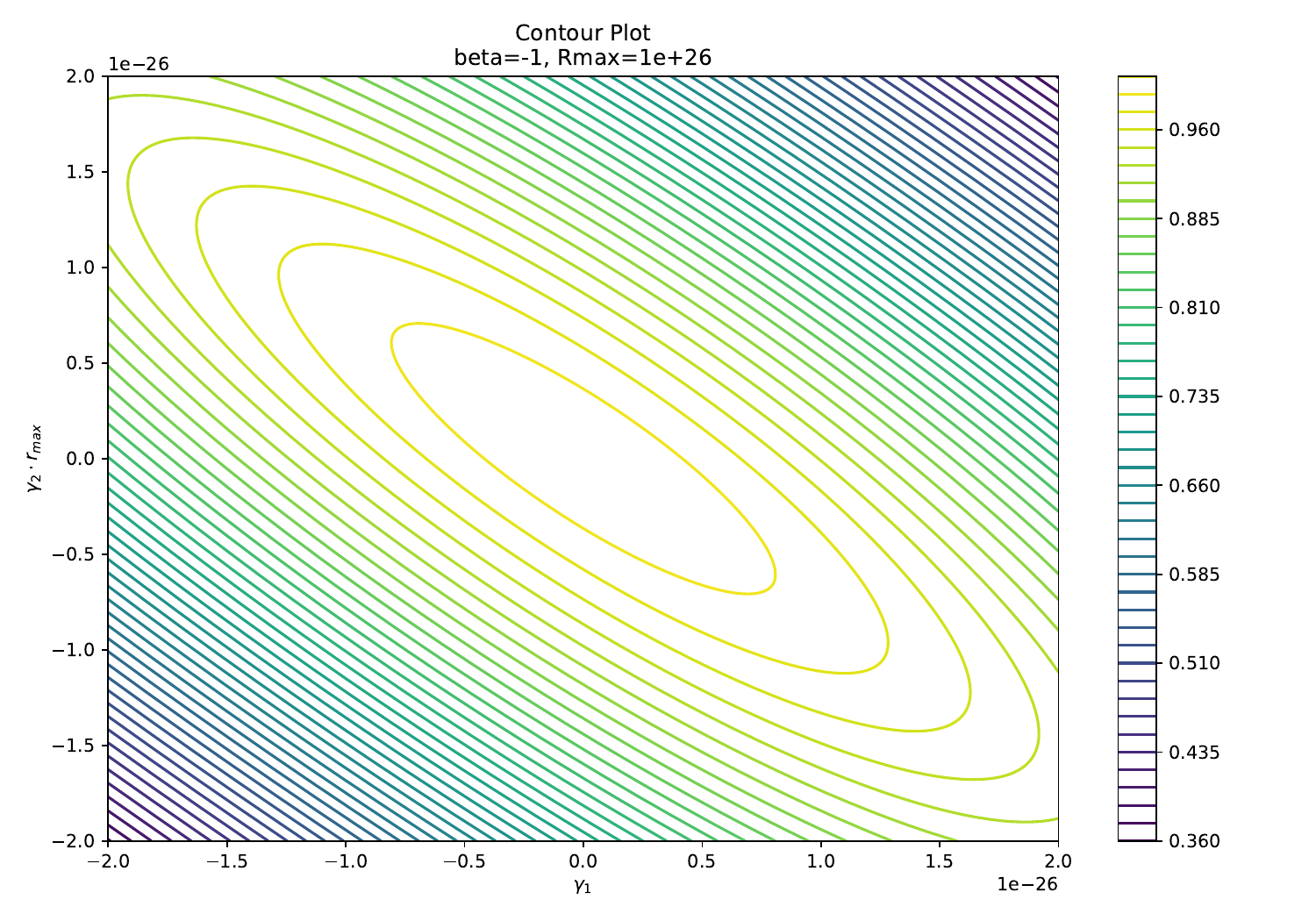}
	\caption{Contour plot of the probability distribution of $\gamma_1$ and $\gamma_2$ defined in Eq.~\eqref{eq: PDF}. The negative correlation of the two variables is easily seen. To enhance the visibility of the plot, we have plotted $\gamma_1$ against $\gamma_2*r_{max}$, and we chose a value of $r_{max}=R_H\approx 10^{26}\,m$, which represents the order of magnitude of the Hubble radius. We picked an indicative value of $\beta=-1$, but different beta will only tune the correlation as long as $\beta<\frac{1}{4}$.}
	\label{fig: contour}
\end{figure}

We see that $\gamma_1$ and $\gamma_2$ are correlated variables with symmetric covariance matrix $\text{Cov}(\gamma_1,\gamma_2)=\Sigma_{ij}$ given by:
\begin{align}
	&\Sigma_{11}=\frac{2  r_{max}^2}{3 (13-36 \beta)}\frac{G_N^2}{ D_{0,T} V},  \nonumber\\
	&\Sigma_{22}= \frac{(5-18\beta)}{9(13-36 \beta ) (1-4 \beta ) }\frac{G_N^2}{ D_{0,T} V}, \nonumber\\
	&\Sigma_{12} = -\frac{ r_{max}}{2 (13-36 \beta)}\frac{G_N^2}{ D_{0,T} V} ,
	\label{eq: CovarianceMatrix}
\end{align}
where $\Sigma_{11}$ and $\Sigma_{22}$ are the variances of $\gamma_1$ and $\gamma_2$ and the two variables have correlation coefficients given by
\begin{equation}
	\rho_{12}=- \frac{3\sqrt{3}}{2\sqrt{2}}\sqrt{\frac{1-4\beta}{5-18\beta}},
	\label{eq: CorrelationCoefficients}
\end{equation}
with the negative correlation easily seen from the plot in Figure~\ref{fig: contour}.
We see that the two variables are negatively correlated with each other and have zero expected value. In other words, if we did not know the value of any of the two constants through other means, we would take their expectation value to be zero. Moreover, we expect one to decrease as the other increases and vice versa.
However, through observational cosmology, we know that our universe presents a positive cosmological constant $\Lambda$, which has to be manually inserted in Einstein's equations by hand and contributes to the Weak field Newtonian limit as
\begin{equation}
	\Phi = -\frac{GM}{r} -\frac{\Lambda}{3}r^2,
\end{equation}
where the $r^2$ dependence represents a global contribution. At this point, we can make the connection with the $\gamma_2$ factor. Given that we observe a value of $\gamma_2=\frac{\Lambda}{3}$, what is the value of $\gamma_1$ that we expect? We can compute this by finding the conditional expectation
\begin{equation}
	\begin{split}
		\mu_{\gamma_1|\gamma_2,r_{max}}&=\mu_{\gamma_1} + \rho_{12}\frac{\sigma_{\gamma_1}}{\sigma_{\gamma_2}}(\gamma_2-\mu_{\gamma_2})\\
		&=-\frac{9}{2}\gamma_2 r_{max}\left(\frac{1-4 \beta }{5-18 \beta}\right),
	\end{split}
\end{equation}
where we know that $\mu_{\gamma_1}=\mu_{\gamma_2}=0$. We now substitute the $\gamma_2=\frac{\Lambda}{3}$ value that we observe and choose $r_{max}=R_H=1.37*10^{26}\,m$ to be the Hubble radius. The Hubble radius gives a scale of the distance beyond which galaxies are receding from us faster than the speed of light due to the expansion of the Universe. Therefore, we arrive at:
\begin{equation}
	\label{eq: Conditional_mean}
	\begin{split}
		\mu_{\gamma_1|\gamma_2,r_{max}}&=-\frac{3}{2}\Lambda R_{H}\left(\frac{1-4 \beta }{5-18 \beta}\right)\\
		&=-2.28*10^{-26}\left(\frac{1-4 \beta }{5-18 \beta}\right).
	\end{split}
\end{equation}

\begin{figure}[H]
	\centering
	\includegraphics[width=0.8\textwidth]{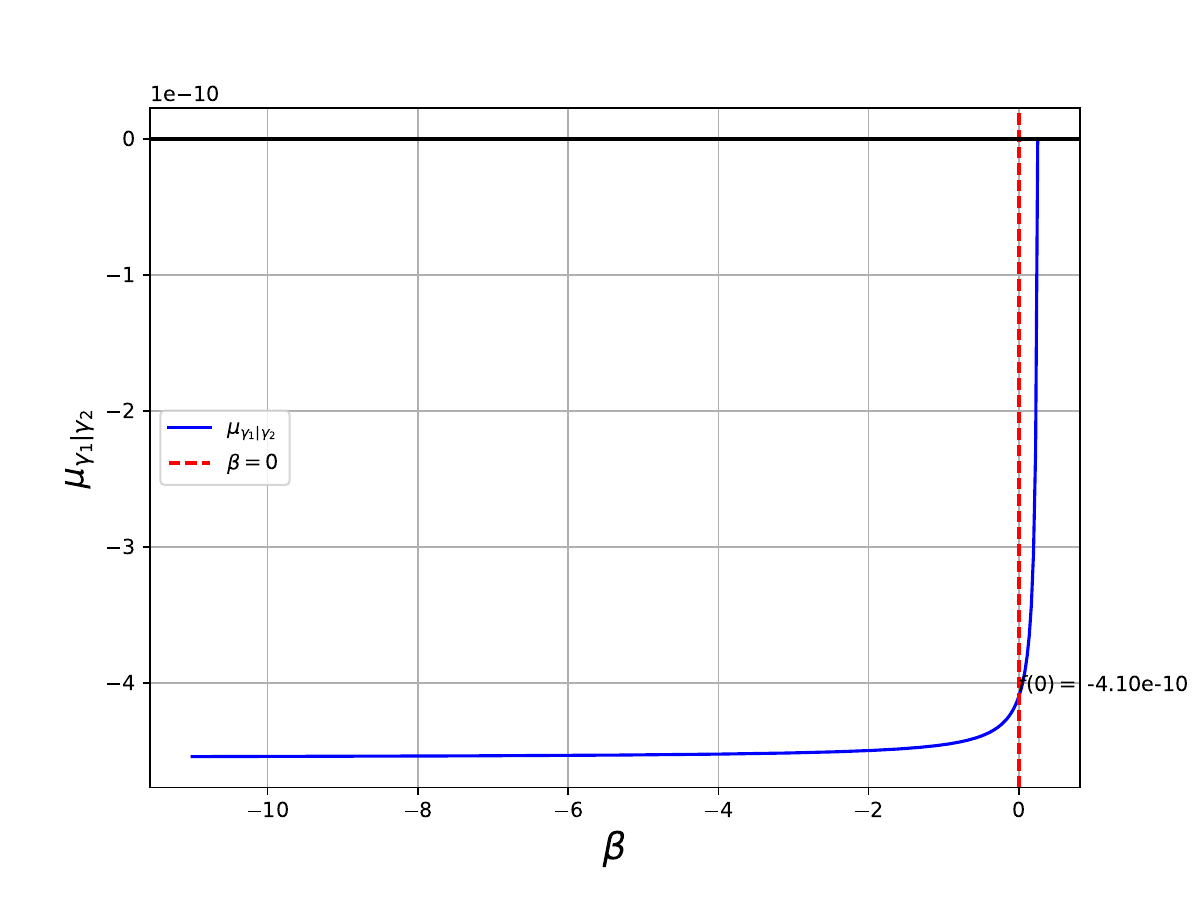}
	\caption{The plot represents the conditional expectation of the value of $\gamma_1$ given the observed value of $\gamma_2$ as a function of $\beta$. For this plot, we have chosen $\gamma_2=\frac{\Lambda}{3}$.}
	\label{fig: expected_gamma1}
\end{figure}

Restoring the units of $c$ means multiplying the above expression by $c^2$, obtaining
\begin{equation}
	\mu_{\gamma_1|\gamma_2,r_{max}}=-2.06*10^{-9} \left(\frac{1-4 \beta }{5-18 \beta}\right),
\end{equation}

which, as $\beta\to 0$ tends to 
\begin{equation}
	\mu_{\gamma_1|\gamma_2,r_{max}}=-4.11*10^{-10} \;m/s^2,
\end{equation}
as plotted in Figure~\ref{fig: expected_gamma1}.

In a dark energy-dominated universe, the value of the Hubble radius can be expressed in terms of the cosmological constant as $R_H=\sqrt{\frac{3}{\Lambda}}$, meaning that the expected value of $\gamma_1$ is
\begin{equation}
	\label{eq: gamma1darkenergy}
	\mu_{\gamma_1|\gamma_2,r_{max}}=-\frac{3}{2}\sqrt{3\Lambda} \left(\frac{1-4 \beta }{5-18 \beta}\right).
\end{equation}

\subsection*{Statistical analysis of the results}
\label{app:stats}
Given the observed value of $\Lambda$, we want to test two things. Firstly, how many standard deviations the observed value of $\gamma_2$ is from the predicted mean of 0, and secondly how many standard deviations the observed value of $\gamma_1$ needed for fitting is. Given that we know the two values follow a bivariate gaussian distribution, we can perform a Z-test. Given that we have a free parameter $D_0$, this would allow us to understand the range of possible values of the decoherence constant required for thee results to sit within 1 standard deviation of their expectation.

To perform the Z-test of $\gamma_2$ we recall that the observed value is $\gamma_2=\frac{\Lambda}{3}$ and compute
\begin{equation}
	Z_{\gamma_2}=\frac{\Lambda / 3}{\sqrt{\Sigma_{22}}},
\end{equation}
where $\Sigma_{22}$ is the variance of $\gamma_2$ and we use the value of the maximal radius as the Hubble radius.
We obtain
\begin{equation}
	\label{eq: Lambda_Ztest}
	Z_{\gamma_2}=\frac{ \Lambda  \sqrt{ D_{0,T} V_H}}{G_N}\sqrt{\frac{(1-4 \beta) (13-36 \beta)}{5-18 \beta }}.
\end{equation}

We want the Z score of $\gamma_2$ to be less than 1, such that the observed value of $\gamma_2$ lies within one standard deviation from the mean. Given that $V_H\approx 10^{79}$, plugging the values in the formula, we obtain
\begin{equation}
	D_{0,T}\leq\frac{Z_{\gamma_2}^2G_N^2}{V_H\Lambda^2}f(\beta),
\end{equation}
with $f(\beta)=\sqrt{\frac{(1-4 \beta) (13-36 \beta)}{5-18 \beta }}$. Dividing by $c^3$ to restore the units, the formula gives
\begin{equation}
	D_{0,T}c^3\leq 1\cdot 10^{-21}f(\beta) \; \frac{m^4}{s\cdot kg^2}.
\end{equation}
Therefore, we see that we are within one standard deviation for any value of $D_{0,T}c^3$ with an order of magnitude less than $10^{-21}f(\beta) \; \frac{m^4}{s\cdot kg^2}$.

We can now perform the same computation for the observed value of $\gamma_1$ This could be the MOND value $\frac{a_0}{c^2}\approx 1.33\cdot 10^{-27}$ or the value of $\gamma$ found in~\cite{mannheim2012fitting}
The conditional variance of the result is given by
\begin{equation}
	\label{eq: Cond_variance}
	\begin{split}
		\sigma^2_{\gamma_1|\gamma_2,r_{max}}&=\Sigma_{11}(1-\rho_{12}^2) \\
		&=\frac{G^2 r_{max}^2}{4  D_{0,T} V(15 -54\beta)},
	\end{split}
\end{equation}
which we can now use to perform the Z-test using the Hubble parameters
\begin{equation}
	\label{eq: gamma1_Ztest}
	\begin{split}
		Z_{\gamma_1} &= \frac{\gamma_{obs}-\mu_{\gamma_1|\gamma_2,r_{max}}}{\sigma_{\gamma_1|\gamma_2,r_{max}}}\\
		&=\frac{2 \sqrt{ D_{0,T} V_H(15-54\beta)}}{G_N R_H} \left(\gamma_{obs}+\frac{3 (1-4\beta) \Lambda  R_H}{2(5-18\beta)}\right),
	\end{split}
\end{equation}
which we can rearrange to obtain
\begin{equation}
	D_{0,T}\leq\frac{G_N^2 R_H^2 Z_{\gamma_1}^2}{V_H}\frac{(5-18\beta) }{3  (2 (5-18\beta) \gamma_{obs}+3 (1-4\beta) \Lambda  R_H)^2}.
\end{equation}
When substituting numbers, restoring units of c and setting $Z_{\gamma_1}=1$, we obtain (for $\beta=-1$)
\begin{equation}
	D_{0,T}c^3 \leq 1.09\cdot 10^{-22} \; \frac{m^4}{s\cdot kg^2},
	\label{eq:D0T-cosmo}
\end{equation}
which means that if $D_{0,T}c^3$ is such that the observed MOND acceleration is within one standard deviation of the conditional expected value, it will automatically be such that the observed value of $\Lambda$ is within one standard deviation of the model.

\section{Comparison with tabletop experiments}
\label{sec:comparison}

In \cite{oppenheim2023gravitationally}, we computed the constraints on temporal fluctuations of the gravitational field, from tabletop precision gravity measurements. These provide constraints on $D_0$, which can be compared with the constraints derived here from galactic rotation curves.
Given the two-point function of the theory, one can compute the variance of the local acceleration. This was done in \cite{oppenheim2023gravitationally} for several two-point functions. The {\it ultra-local, non-relativistic theory}, has a two-point function for $\Phi(x)$ in the weak field limit, given by
\begin{align}
	G_2(x,x')=\frac{G_N^2}{(4\pi)^2(1-\beta)D_{0}}\int\frac{ d^3y \delta(t,t')}{|\vec{y}-\vec{x}||\vec{y}-\vec{x}'|}\quad .
	\label{eq:nofreedom}
\end{align}
This corresponds to a model which is continuous in phase space, and corresponds to the linearised weak field limit of GR with local white noise. This model was found to be ruled out by experiment, since the variance in the potential divergences. This is effectively an IR divergence since $\langle\Phi(x)\Phi(x')\rangle$ diverges over finite distances with the volume of the ambient space unless $D_2$ depends strongly on the potential.
However, the two point-function of \eqref{eq:nofreedom} is related to the one given in~\cite{grudka2023renormalisable}, namely
\begin{align}
	G(x,x'):
	=&-\frac{G_N^2}{8 \pi (1-\beta)D_0}|\vec{x}-\vec{x}'|\delta(t,t').
	\label{eq:freedom}
\end{align}
One can see the relationship between Eq.~\eqref{eq:nofreedom} and Eq.~\eqref{eq:freedom} by using dimensional regularisation to perform the integral in Eq.~\eqref{eq:nofreedom}~\cite{morris2024dim}. 
While Eq. \eqref{eq:freedom} is not postive semidefinite, for any finite range of $x,x'$ one can add a large enough constant to insures that~\eqref{eq:freedom} is positive semi-definite. The constant does not effect the variance in acceleration which is the physically meaningful quantity. We conjecture that this is thus a gauge choice, but a fuller understanding of the physical implications is required. 
These two-point functions can be found from the non-relativistic limit of the two-point function for the trace of the scalar mode of the full relativistic theory, when the bare cosmological constant is taken to be zero.

We can use the regularised two-point function to place bounds on $D_0$ from tabletop experiments, and we will find that the theory is consistent with experiment, indicating that the IR divergence appears to be an artifact of the unregularised theory, leading to the conclusion that the ultra-local theory is not ruled out by experiment even when linearised.
The acceleration covariance matrix is obtained from Eq. \eqref{eq:freedom}
\begin{align}
	\frac{\partial^2 G(x,x')}{ \partial{x_i}\partial{x'_j}}=\frac{G_N^2}{8\pi (1-\beta) D_{0}} \left( \frac{\delta_{ij} |x - x'|^2 - (x_i - x'_i)(x_j - x'_j)}{|x - x'|^3} \right)\delta(t,t')
\end{align}
and by setting $i=j$ and summing, we obtain the variance of the acceleration $\sigma_a^2(x,x'):=\langle\nabla\Phi(x)\cdot\nabla\Phi(x')\rangle-\langle\nabla\Phi(x)\rangle\cdot\langle\nabla\Phi(x')\rangle$
\begin{align}
	\sigma_a^2(x,x')=\frac{G_N^2}{4\pi (1-\beta)D_{0}}\frac{1}{|\vec{x}-\vec{x}'|}\delta(t,t')
	\label{eq:AccVartt}
\end{align}
In equilibrium, when the sources only change slowly, the relativistic covariance matrix is obtained by replacing $\delta(t,t')$ by $\delta(t- |x-x'|,t')$, corresponding to the retarded propagator. 

Although this quantity diverges as $x\rightarrow x'$, it is finite when smeared over test masses, which is the physically meaningful quantity since we can only measure the acceleration of some test mass. 
For test masses with total mass M, and mass densities $m(x)$, we can consider
the force they feel 
\begin{equation}
	\bar{F}:=\frac{1}{2T}\int_{-T}^T d^3x dt\, m(x)\nabla\Phi(x)
\end{equation}
averaged over some time $2T$. The variance 
of this averaged force is then
\begin{align}
	\bar{\sigma}_F^2:=\frac{1}{(2T)^2}\int d^4x d^4x' m(x)m(x')\sigma^2_a(x,x')
	\label{eq:acceleration_variance}
\end{align}
integrated over the test mass and time $2T$ the duration of the experiment.

The average variance of the corresponding acceleration is thus
$  \bar{\sigma}_a^2$ is 
\begin{align}
	\bar{\sigma}_a^2:=&\frac{1}{2T(M)^2}
	\frac{G_N^2}{4\pi D_{0}(1-\beta)}\int d^3x d^3x' \frac{m(x)m(x')}{|\vec{x}-\vec{x}'|}
\end{align}
where we have divided by the total mass $M^2$ to convert from a variance of force to a variance of acceleration. The quantity  
\begin{align}
	U=\int d^3x d^3x' \frac{m(x)m(x')}{|\vec{x}-\vec{x}'|}
\end{align}
is seen to be akin to the Newtonian gravitational self-energy of the system. For a uniform density sphere of radius $R$, this is $3M^2/5R$, giving
\begin{align}
	T \bar{\sigma}_a^2= \frac{3G_N^2}{40\pi D_{0}c^3(1-\beta) R}
	\label{eq:SphereSpec}
\end{align}
where we have put back factors of $c$. Converting to angular acceleration which is often more relevant for torsion pendulums would bring a factor of $R^{-2}$. 

One can often obtain a more precise measurement of acceleration by monitoring the acceleration $a(t)$ over time, and then Fourier transforming it to obtain  $\hat{a}(\omega)$. The variance in the acceleration at time $t$ is given by integrating Eq. \eqref{eq:AccVartt} over $t'$. The Fourier transform of this variance at time $t$,  is the spectral density $S_{aa}(\vec{x},\vec{x}';\omega)$. In the weak field limit, this matches the relativistic result computed in~\cite{grudka2023renormalisable} for the retarded and advanced propagator. In the massless case, for the retarded and advanced propagator, it is found to be
\begin{align}
	S_{aa}(\vec{x},\vec{x}';\omega):=\frac{G_N^2}{ 4\pi (1-\beta)D_0}\left(\frac{1}{|\vec{x}-\vec{x}'|}\pm \frac{i \omega }{2}\right) e^{\pm i \omega |\vec{x}-\vec{x}'|}
	\label{eq:spectral}
\end{align}
where we have fixed the prefactor such that this corresponds to the action of Eq.~\eqref{eq:main-isotropic_zero_order} in the weak field limit. The second term is not positive semidefinite in the case of the pure retarded propagator without imposing additional conditions, but this is not relevant to the stationary case~\cite{grudka2023renormalisable}. One can see this by taking the Fourier transform of Eq.~\eqref{eq:spectral}, which result in a $\delta(t-t'\pm |x-x|)/|x-x'|$ for the first term and a 
$\delta'(t-t'\pm |x-x|)/2$ for the second.
At low frequency, this spectral density integrated over a test mass $S_{aa}(\omega)$ is given by the right hand side of Eq.~\eqref{eq:SphereSpec}. On the other hand, at very high frequencies, the oscillations from the exponential term act as an effective cut-off, since we have
\begin{align}
	\int_{-\omega_c}^{\omega_c}d\omega S_{aa}(\vec{x},\vec{x}';\omega)=\frac{G_N^2}{ 2\pi (1-\beta)D_0}\left( \frac{1}{|x-x'|^2} \pm \frac{i \omega_c }{2|x-x'|} \right) \sin { \omega_c |\vec{x}-\vec{x}'|}
\end{align}
and
\begin{align}
	\lim\limits_{\omega_c\rightarrow\infty}\frac{\sin{\omega_c |x-x'|}}{2\pi |x-x'|}=8\pi \delta^{(3)}(\vec{x},\vec{x}')|\vec{x}-\vec{x}'|^2
\end{align}

The low frequency regime is relevant for most experiments, in which case Eq.~\eqref{eq:SphereSpec} closely tracks the spectral density. 
Experimental upper bounds on the relevant spectral density have been collected in~\cite{Carney_2021,janse2024current}. Here, we exclude differential acceleration measurements such as those found in~\cite{asenbaum2017phase,armano2018beyond}, since stochastic fluctuations at low frequency may affect both masses or paths in the same way, and requires some further understanding of on-shell vs off-shell contributions.
The experiment of \cite{hoyle2004submillimeter} uses a torsion pendulum with a moment of inertia $I\sim 10^{-7} \unit{kg.m^2}$ and finds a torque variance of order $10^{-27}\unit{(Nm)^2/Hz}$ at the $\unit{mHz}$ 
scale. The disk radius is of order $\unit{cm}$, leading to $S_{aa}(\omega)\sim 10^{-17}\unit{(m/s^2)^2/Hz}$ (since we are only estimating orders of magnitude we can use \eqref{eq:SphereSpec} for spherical test masses, since  $U=2M^2/3R$ for the disk 
geometry of \cite{hoyle2004submillimeter}) is of the same order of magnitude as for the sphere. Taking $1-\beta$ to be of order unity leads to the dimensionless coupling constant being $G_N^2/
D_0c^6\leq 10^{-42}$ ($D_0c^3\geq 10^{-2} \unit{ m^3/kg^2.s}$). The experiment of \cite{westphal2020measurement} uses $\unit{mm}$ 
size masses, and reports $S_{aa}(\omega)\sim 10^{-18}\unit{(m/s^2)^2/Hz}$, leading to 
$G_N^2/D_0c^6\leq 10^{-43}$ ($D_0 c^3\geq 10^{-1}\, \unit{m^3/kg^2.s}$) also at the $\unit{mHz}$ scale. 
In terms of future experiments, LISA Pathfinder found $S_{aa}\sim 10^{-30}\unit{(m/s^2)^2/Hz}$ differentially, using $\unit{cm}$ scale test masses, so once LISA becomes operational, it should be able to improve the bound by at least $10$ orders of magnitude. The bound depends on $U$ (generally $\sim 1/R$) but not the mass, so using heavy test masses of small size will generally give better bounds.

At present, these lower bounds on $D_0$ are consistent with current upper bounds due to interference experiments and the decoherence vs diffusion trade-off.   
The decoherence rate for the ultra-local non-relatistic model is given by~\cite{oppenheim2023gravitationally} $\lambda=2D_{0}c^3M^2/V_\lambda$, where $M$ is the mass of the particle in the interference experiment, and $V_\lambda$ is the volume of the wave-packet. For $M\approx 10^{-24}$ kg for fullerene molecules, $V_\lambda\approx 10^{-25}\unit{m^3}$ for the wavepacket volume estimated in the experiment of~\cite{Gerlich2011} and a decoherence rate of  $\lambda\geq 0.1 \unit{s^{-1}}$~\cite{Gerlich2011},
gives $D_0c^3\leq 10^{24} \unit{m^3/kg^2 s}$, and so we have a decoherence vs diffusion squeeze of
$10^{-2}\leq D_0c^3\leq 10^{24} $. That this upper bound could be lowered to match the lower bound which could be obtained by LISA appears feasible.

Let us now consider secondary decoherence~\cite{tilloy2017principle}, that is the additional decoherence caused by the fact that the stochastic contribution to the matter Hamiltonian, causes additional decoherence in addition to that given by the Lindblad operators. Since the mass term of the Hamiltonian (e.g. $\frac{1}{2} m^2 \int \sqrt{g}\phi^2$ for a massive scalar field $\phi$) depends on the metric which is fluctuating, it will cause further decoherence. For the two-point function of Eq. \eqref{eq:freedom}, and separated mass distributions this has been calculated 
to be~\cite{oppenheim2024secondary}
\begin{align}
	\lambda=2D_0 c^3 \frac{M^2}{V_\lambda}+\frac{G_N^2}{D_0 c^3}\frac{M^2}{8\pi \hbar^2}|x_L-x_R|
	\label{eq:secondary}
\end{align}
where $|x_L-x_R|$ is the spatial separation of the superposition, which can be taken to be the spacing of the diffraction grating (typically $\sim 100 nm$). We have also included the primary decoherence term in Eq \eqref{eq:secondary}. The lower bound on $D_0 c^3$ of $10^{-2} \unit{m^3/kg^2.s}$ imply that the secondary decoherence is negligible at $\leq 10^{-7}\unit{s^{-1}}$.

\section{The stochastic action for the isotropic metric, and the Newtonian limit}
\label{app: isotropic}

For the purpose of this section, we only consider a static matter distribution with negligible contributions from matter pressure, frame velocity and specific energy density. In other words, we are only interested in higher-order corrections coming from the gravitational potential $\Phi$ itself. We implicitly choose a homogeneous isotropic universe in which resides an isolated Post-Newtonian system with coordinates such that the outer region far from the isolated system is in freefall with respect to the surrounding cosmological model but at rest with respect to a frame in which the universe appears isotropic. It is then possible to show that one can construct a local quasi-Cartesian system in which metric and matter degrees of freedom can all be evaluated consistently with the Post-Newtonian approximation. Lastly, one might need to take into account the extent of preferred frame effects including frame dragging and the coordinate velocity of the frame relative to the mean rest frame of the universe. All the aforementioned effects can be summarised through what is known as the Parametrised Post-Newtonian formalism (PPN), whose first formulation dates back to Eddington in 1922. When formulated in a coordinate frame moving along with the physical system of interest, post-Newtonian effects can be summarised through the metric (with units of c):
\begin{equation}
	\label{eq: PPN_metric}
	\begin{split}
		&g_{00}\approx-c^2\left(1+\frac{2\Phi}{c^2}+\frac{2\beta\Phi^2}{c^4}+f(\alpha_i,\beta,\gamma,\zeta_i,V_i,W_i)\right)+\mathcal{O}(c^6), \\
		&g_{ij}\approx\left(1-\frac{2\gamma\Phi}{c^2}\right)\delta_{ij}+\mathcal{O}(c^4),\\
		&g_{0i}\approx h(\alpha_i,\gamma,\zeta_i,V_i,W_i)+\mathcal{O}(c^{5}),
	\end{split}        
\end{equation}
where $\alpha_i$, $\zeta_i$ with $i=\{1,2,3\}$ represents respectively the extent of preferred frame effects and the extent of failure in the conservation of energy, $\beta$ measures the amount of nonlinearity in the superposition law for gravity, $\gamma$ the amount of curvature produced by a unit rest mass and $V_i,W_i$ effects related to the frame velocity \cite{PPN,Will:2011nz}. The strength of the Parametrised Post Newtonian formalism is that it can be applied to theories of gravity outside of general relativity. However, to describe the post-Newtonian limit of general relativity one takes $\alpha_i=\zeta_i=0$ and $\beta=\gamma=1$, which is what we will do in this paper.

Given these premises, we write the isotropic metric as
\begin{equation}
	ds^2=-c^2 e^{\frac{2\Phi}{c^2}}dt^2+e^{-\frac{2\Phi}{c^2}}\delta_{ij}dx^idx^j.
\end{equation}
One may worry that the exponential form of this metric may not be consistent at higher orders in the expansion, for example, not all terms in the expansion may be physically relevant. However, for all effects and purposes, in this paper, we will never exceed order $\mathcal{O}(c^{4})$, such that the metric matches perfectly with the PPN formalism.
For the matter distribution, we will take the Stress-Energy tensor to be that of pressureless dust, being given by
\begin{equation}
	T_{00}=m\, e^{2\Phi}, \quad T^{ij}=0, \quad T^{0i}=0.
\end{equation}

Using the isotropic metric (with $c=1$), the components of the action of Eq. \eqref{eq: actionG}
\begin{align}
	&G^{\mu\nu}G_{\mu\nu}= e^{4\Phi} \left( 3\left(\nabla\Phi\right)^4+\left( \left(\nabla\Phi\right)^2-2
	\left(\nabla^2\Phi\right)\right)^2\right),\\
	&G^2=4e^{4 \Phi} \left(  \left(\nabla\Phi\right)^2-\nabla^2\Phi\right)^2,
\end{align}
The coupling to matter can be deduced from the final line of Eq \eqref{eq: CQaction} since in the classical limit the system decoheres and for a decohered system, there is no distinction between
$\bar{T}_{\mu\nu}$ and $T_{\mu\nu}$.
The components are given by
\begin{align}
	G^{\mu\nu}T_{\mu\nu} &= -m\,e^{2\Phi}\left(  \left(\nabla\Phi\right)^2-2\left(\nabla^2\Phi\right)\right),\\
	T^{\mu\nu}T_{\mu\nu}&=m^2,\\
	T^\mu_\mu&=-m.
\end{align}
where $T$ is the trace of the stress energy tensor.

The full action for the isotropic metric is thus
\begin{equation}
	\actionC=-\frac{D_0 c^5}{64\pi^2G_N^2}\int d^3\vec{x}dt\; e^{\frac{2\Phi}{c^2}}\left[\left(\nabla^2\Phi-\frac{(\nabla\Phi)^2}{2c^2}-4e^{-\frac{2\Phi}{c^2}}\pi Gm\right)^2+\frac{3}{c^4}(\nabla\Phi)^4-4\beta\left(\nabla^2\Phi-\frac{(\nabla\Phi)^2}{2c^2}-4e^{-\frac{2\Phi}{c^2}}\pi Gm \right)^2\right]
	\label{eq: isotropic action}
\end{equation}
where we have put in powers of $c$ as one can use it to perform an expansion in powers of $1/c^2$. One immediately sees that at $0'th$ order in $1/c^2$, we recover the Newtonian action of \eqref{eq:main-isotropic_zero_order}.

\section{Entropic forces}
\label{sec:Brownian}

A canonical example of an entropic force is that due to a polymer which is initially curled up in a low entropy state, but will unfurl or diffuse into a higher entropy state, with its ends exerting a force~\cite{neumann1977entropy,roos2014entropic}. Another is a gas in a box fitted with a piston on one side, which is slowly pushed out as the gas diffuses.
Note that in the main body of this article, we do not consider deriving gravity as an entropic force~
\cite{jacobs2014open,padmanabhan2010thermodynamical,verlinde2011origin,visser2011conservative,van2010building}, but rather consider the entropic force that gravity exerts.
The purpose of this section is to define entropic forces as applicable out of equilibrium and based only on the equations of motion. It will also give an example that can be solved in a similar manner to the gravitational case and has similar features.

Consider Newton's law $F(q)=m\ddot{q}$. This is a deterministic equation, but we can consider the case where the system is in a probability distribution over $q$, in which case we still expect Newton's law to be satisfied on expectation
\begin{align}
	\langle m\ddot{q} - F(q)\rangle=0.
\end{align}
The important ingredient is that the mean value of the force felt by the particle depends on the second and higher moment of its position, and so it generally doesn't follow its deterministic trajectory because the average of the position equation of motion is not the same as the equation of motion of the average position. We therefore define the entropic force $F_S$ to be
\begin{align}
	F_S(q)=F(\langle q \rangle)-\langle F(q) \rangle,
\end{align}
since it captures the extra force due to diffusion.
A simple example is given by $F=-\alpha q^2$ for the  cubic potential corresponding to $V(q)=\alpha q^3/3$. The time derivative of the particle's mean momentum obeys $\langle\dot{p}\rangle=\alpha\langle q^2\rangle$ which can be significantly larger than $\langle q\rangle^2$. Another example is Brownian motion of a particle in a box with a piston.
The presence of a wall on the other side suffices to ensure that the mean value of the particle's position $q$ will change with time as the piston is pushed out. If there were no diffusion or wall, the particle's average position does not change. The wall placed at $q=0$ makes it impossible for $\langle q_f\rangle=q_0$ when $\langle q^2\rangle$ is non-zero. After all, given enough time, the reflecting boundary at the origin will skew the average final position in the direction opposite to the wall. 

Indeed, as $\langle q^2\rangle$ becomes greater and greater than $\langle q\rangle^2$, (possibly due to elapsed time or a temperature increase of the heath bath) the presence of the wall makes it so that the average final position will be further and further away from the mean. We will call this the \textit{diffusion regime}, since the second moment of the observable is comparable to its variance and is influencing the observable equations of motion, in comparison to the case where the mean value of the observable is given by its deterministic value, which for a free Brownian particle corresponds to the final position being identical to its initial position $\langle q_f\rangle\approx q_0$.

We will now explicitly show the example of a Brownian particle with a wall, and show that it has very similar features to the gravitational example discussed in the main body of the paper, and see that it can be solved in a similar way.

\subsection*{Brownian motion with a wall}

Consider the path integral for a free particle undergoing Brownian motion with no drift. The probability of finding a particle at $q(t_f)=q_f$ given that at $t=0$ it was at $q(0)=q_0$ and had velocity $v_0$ and acceleration $a_0$ is given by the Onsager-Machlup path integral
\begin{align}
	P(q_f|q_0,\dot{q}_0,\ddot{q}_0)=\frac{1}{\mathcal{N}}\int_{q_0}^{q_f} \mathcal{D}q\, e^{-\frac{1}{2 D_2}\int_{0}^{t_f} (\ddot{q})^2 dt}.
	\label{eq:PIbrownian}
\end{align}
Note that the path integral acts to suppress the probability of paths which do not satisfy $\ddot{q}=0$, by an amount controlled by the diffusion constant $D_2$. The larger $D_2$ is, the more stochasticity we are likely to find in the paths which are realised.
This is an equivalent description of the dynamics often described by the Langevin equation, $\ddot{q}=F(q)/m+j(t)$, with $F/m$ the drift produced by a deterministic force $F$ (here set to $0$), and $j(t)$ a stochastic white noise process. The dynamics can also be described via the Fokker-Planck equation~\cite{risken1996fokker}  or Ito calculus~\cite{gardiner2004handbook} and we refer the interested reader to~\cite{risken1996fokker} for a derivation of the Onsager-Machlup path integral, or \cite{feynman1965quantum} for a discussion of Brownian motion in the context of path integrals of similar form to Eq. \eqref{eq:PIbrownian}.

We now imagine that there is a step function $V\Theta(-q)$ potential (we could take $V\rightarrow\infty$). This prevents the particle from going to the negative values. we can express this by modifying the OM Lagrangian to be
\begin{equation}
	\mathcal{L}_{OM}(\ddot{q})= -\frac{1}{2D_2}\left(\frac{d^2}{dt^2}|q|\right)^2,
\end{equation}
the variation of the Lagrangian provides the fourth-order Euler-Lagrange equation for the most probable paths:
\begin{equation}
	\frac{d^4}{dt^4}|q|=0,
\end{equation}
with general solution 
\begin{equation}
	q_{MPP}(t)=\alpha_0 + \alpha_1 t+\frac{1}{2}\alpha_2 t^2 + \frac{1}{6}\alpha_3 t^3.
	\label{eq:Qmpp}
\end{equation}
This is remarkably similar to~\eqref{eq:OurAction}.
When substituting back into the action, we see that the terms corresponding to the deterministic solution $\alpha_0$ and $\alpha_1$ (which is the global minimum) drop out due to the second-order time derivative. Therefore, we are allowed to fix them through initial conditions on $q(0)$ and $\dot{q}(0)$. The action then takes the form of a bivariate Gaussian distribution, which when integrated from the initial time $t_0=0$ to the final time $t_f$ becomes:
\begin{equation}
	e^{\mathcal{S}_{OM}}=\exp\left(-\frac{t_f}{3 D_2} \left(3 \alpha _2^2+3 \alpha _3 \alpha _2 t_f+\alpha _3^2 t_f^2\right)\right).\label{eq:Qmppaction}
\end{equation}

At this point, we can relate $\alpha_2$ and $\alpha_3$ to other known initial conditions or final conditions, and the action will act as the probability weight of the most probable path given the specified conditions. However, we could also use it to find the average final position. For the sake of simplicity and to obtain an analytical expression, we assume that the particle starts with no acceleration and that at the final time is at positon $Q$.
In particular, we fix $q(0)=q_0>0$ and $\dot{q}(0)=\ddot{q}(0)=0$, such that the particle begins on the right-hand side of the wall with zero initial velocity and acceleration. This fixes $\alpha_1=\alpha_2=0$. The last condition is fixed by setting $q(t_f)=Q\geq 0$, the final position of the particle, arriving at
\begin{equation}
	q(t) = q_0 + \frac{(Q-q_0)t^3}{t_f^3}.
\end{equation}
We can now substitute the solution into the Lagrangian to perform a saddle point approximation and integrate it up to the final time. We arrive at the action which determines the probability weighting of the most probable path given initial and final conditions:
\begin{equation}
	\mathcal{S}_{OM}(Q)= -\frac{6(Q-q_0)^2}{D_2 t_f^3},
\end{equation}
we can now integrate over all possible final positions to normalise the integral
\begin{equation}
	\int_{-\infty}^{\infty} dQ\, P(Q|q_0,\dot{q}_0,\ddot{q}_0)=\frac{1}{\mathcal{N}}\int_{-\infty}^{\infty} dQ\, e^{-\frac{6(Q-q_0)^2}{D_2 t_f^3}} = 1,
\end{equation}
to find 
\begin{equation}
	\mathcal{N} = \sqrt{\frac{D_2\pi\, t_f^3}{6}}.
\end{equation}
At this point, we can compute the average final position by keeping in mind that there is a wall at $q=0$ such that
\begin{equation}
	\label{eq: final_avg}
	\begin{split}
		\langle Q\rangle &= \sqrt{\frac{6}{D_2\pi\, t_f^3}}\int_{-\infty}^{\infty} dQ\, |Q| e^{-\frac{6(Q-q_0)^2}{D_2 t_f^3}} \\
		&= \frac{1}{12} \left[6q_0+\left(1+\Gamma_{\mathcal{R}}\left(-\frac{1}{2},0,\frac{6 q_0^2}{D_2 t_f^3}\right)\right)+\sqrt{\frac{6D_2}{\pi }} t_f^{3/2} e^{-\frac{6 q_0^2}{D_2 t_f^3}}-6 q_0 \text{Erf}\left(\sqrt{\frac{6}{D_2 t_f^3}}q_0\right)\right],
	\end{split}
\end{equation}
where $\text{Erf}$ is the error function and $\Gamma_{\mathcal{R}}$ is the regularised Gamma function. 

\begin{figure}[H]
	\centering
	\includegraphics[width=0.9\textwidth]{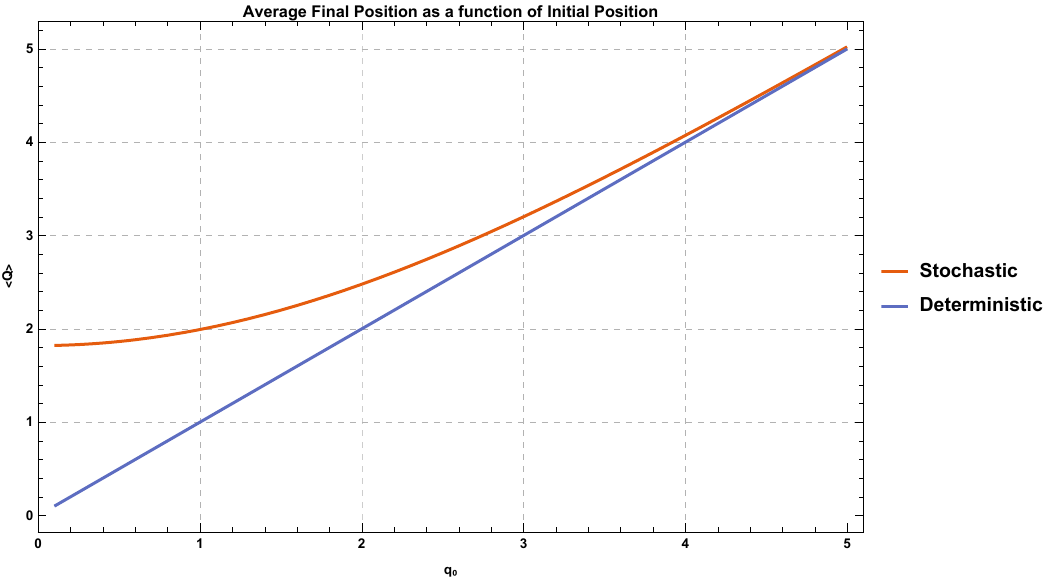}
	\caption{Average final position as a function of initial position according to Equation~\eqref{eq: final_avg} for fixed final time $t_f=5$ and diffusion coefficient $D_2=\frac{1}{2}$. In the presence of a wall, the closer the Brownian particle starts to the reflective wall at $q=0$, the more its average final position will diverge from its deterministic value. The particle is assumed to start with zero velocity and acceleration.}
	\label{fig: avg_final_pos}
\end{figure}

This solution is very insightful. As the diffusion vanishes $D_2\to 0$ or the final time goes to zero $t_f\to 0$, the argument of the error function, the exponential and the regularised gamma function go to infinity. The error function and the exponential vanish while the gamma function becomes 1, leaving $\langle Q\rangle = q_0$. As one would expect for a situation where there is either no diffusion or no time has elapsed, the final average position is the same as the initial one, which is also the deterministic behaviour.
Even more interesting, the same happens when $q_0$ is very large; indeed, if the particle is very far from the wall, it will not feel its effect until enough time has passed, as it can be seen in Figure~\ref{fig: avg_final_pos}, such that there is an opposite effect between the growth of $q_0$ and that of $t_f$.

\begin{figure}
	\centering
	\includegraphics[width=1.1\textwidth]{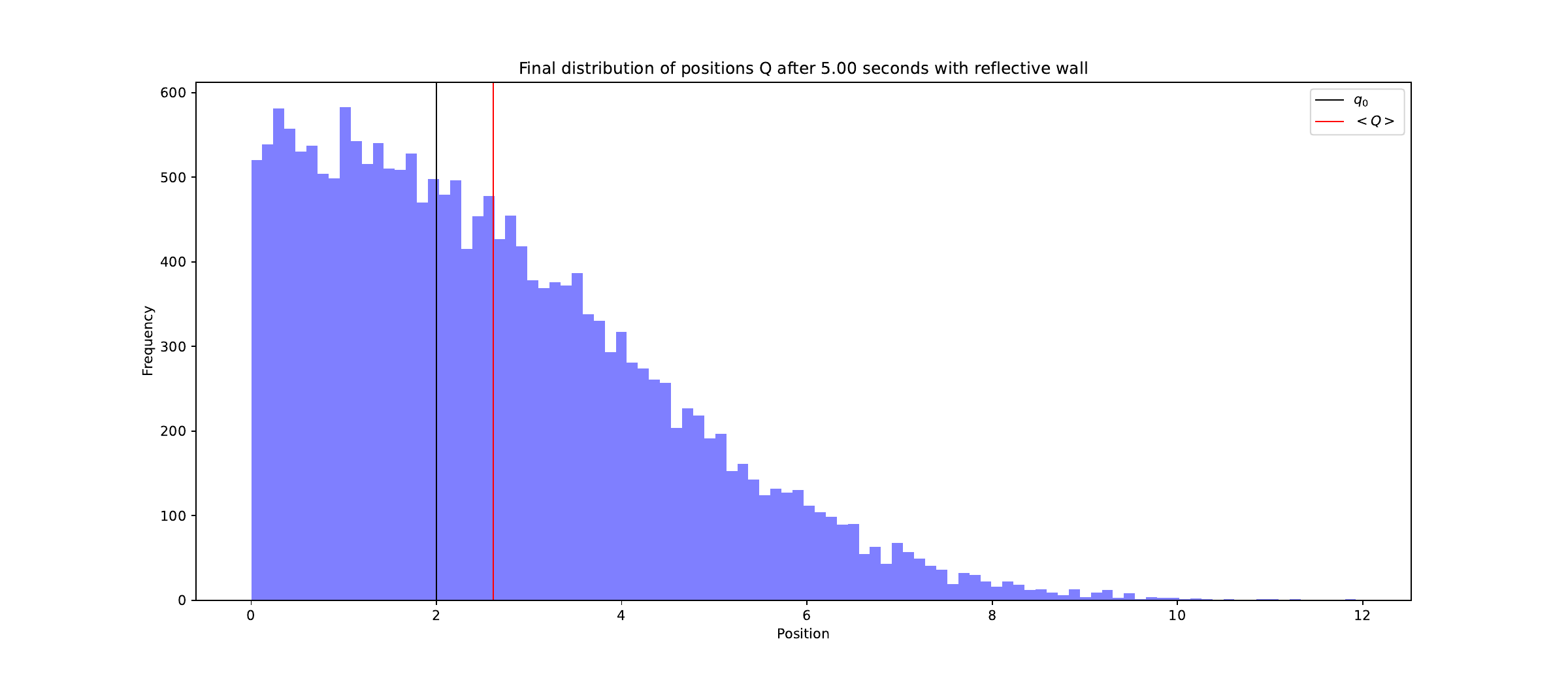}
	\caption{Monte Carlo simulation of the probability density function of the final position for the Brownian particle with a wall with $t_f=5$. As one can see, the initial position is $q_0=2$. In the deterministic case, this should correspond to the final position in the absence of initial velocity and acceleration. However, the average final position is skewed to the right, as the presence of the wall produces an entropic force which creates a deviation.}
	\label{fig: brownianwall}
\end{figure}

Lastly, one could assume the particle is not too far from the wall and perform a short time expansion to arrive at
\begin{equation}
	\langle Q\rangle = q_0 + \frac{1}{2}\sqrt{\frac{D_2}{6\pi} t_f^{3}},
\end{equation}
such that one sees that the average final position increases as $t^{3/2}$. In Fig.~\ref{fig: brownianwall} we show the probability density function of the final positions obtained from a Monte Carlo simulation.

\section{Comment on Analysis of anomalous contribution to galactic rotation curves due to stochastic spacetime}
\label{sec:comment}

{\bf\noindent Comment 1:} Two parts of v1 of this preprint were criticized in~\cite{hertzberg2024comment}, neither of which formed part of our analysis. 
\\\\
{\bf\noindent Is the $\gamma_1 r$ term ''forbidden''?:} First, the authors of~\cite{hertzberg2024comment} note that if in the Newtonian case, one has a $\kappa_1 r$ term in the potential in the vacuum, then it  cannot satisfy matching boundary conditions for a localised source if it is to satisfy the MPP equation with matter $\nabla^4\Phi(x)=4\pi G_N\nabla^2m(x)$. This is true. However, the authors of ~\cite{hertzberg2024comment} then conflate the MPP equation for an equation of motion, distinguishing what they call ''vacuum solutions'' in comparison to fluctuations. They claim that the $\kappa_1 r$ term is ''forbidden''. It is not. As already emphasised in v1, the MPP is not an equation of motion but only gives {\it the most probable path}. These paths are themselves fluctuations, and other paths including the $\kappa_1 r$ term in vacuum contribute. This can be seen by also inserting the $\kappa_1$ term into the action which gives in the spherically symmetric case
\begin{align}
	I_{MPP}'=-\frac{D_{0,T} (1-\beta)}{G_N^2}\int r^2 dr
	\left(6\kappa_2
	+\frac{2\kappa_1}{|r|}
	\right)^2\quad.
	\label{eq:NewtMPPaction2}
\end{align}

We see $\kappa_1$ anti-correlated with the cosmological constant term $\kappa_2$. While suppressed, it could still contribute enough to influence galactic rotation curves.
The path integral calculation demonstrates this. If you observe an $r^2$ term (which is an extremal path) corresponding to a fluctuation which acts as a cosmological constant then you should expect to see an $r$ term.  This is unchanged when a localised matter distribution is present. In the actual calculation, we consider a power series expansion of the fluctuations. This is reasonable since any function can be expanded as a power series, and contrary to what is claimed in \cite{hertzberg2024comment}, we do consider all terms in the power series. It is true however, that a fuller principle component analysis would be useful, via the Kosambi–Karhunen–Loève theorem~\cite{Love1978}, to determine the appropriate contributions more fully. 

{\bf\noindent MOND behavior up to a constant:}  Next, they claim that in our v1 preprint, we drop a constant term in our discussion, but we did not, we explicitly keep and note it.  This was in the context of a discussion, where we tried to give the reader further intuition concerning Eq. \eqref{eq: NewtVar}, which very clearly does set a natural acceleration scale. When the mean acceleration drops below its variance, the acceleration will act on expectation as a positive matter source. The separate discussion around the different anomalous contributions to rotation curves now comes after Eq. \eqref{eq: Conditional_mean-main} which we have expanded. The linear term alone provides a reasonable fit for larger spiral galaxies~\cite{mannheim1989exact},  but it is true that to account for more recent data, $\gamma_2$, is taken to be a $\kappa \approx 10^{-50} m^{-2}$~\cite{mannheim2012fitting} to ''flatten the curve''. This was noted in v1 of this manuscript.

They make a third point already discussed by us in v1 
-- that by itself, the random variables found in Eqs.~\eqref{eq:NewtMPPaction2} or \eqref{eq:OurAction} are zero on expectation. This is true, but because they have variance, we would be very surprised if we observed a cosmological constant of zero. Furthermore, given that the galaxies we observe all formed under the dynamics of the same cosmological constant, we would also expect each of their linear fluctuations to be anti-correlated with the cosmological constant term, provided the conditional variance in the linear term is small enough, which indeed is the case. If we post-select on one global observation, it can provide a boundary condition for the remaining terms. This however does depend on the underlying dynamics in cosmology,
a point made often in v1.

Furthermore, although in the linearised theory, there is no preferred sign to the terms, this is not the case once non-linear corrections are included, as \eqref{eq: NewtVar} makes clear, and in the example of Brownian motion with a potential given in Appendix \ref{sec:Brownian}. We find evidence for this in \cite{UCLcosmo}.
\newline

{\bf\noindent Comment 2:} In v2-v4 of ~\cite{hertzberg2024comment}, the authors raised several additional points.

{\bf\noindent Comparison of the Schwarschild power spectrum with the CMB power spectrum:} The authors compare the power spectrum of stochastic fluctuations in the stochastic Schwarszchild spacetime, to the dark matter contribution to the CMB power spectrum in the cosmological setting and note that they are different.
However, this comparison is not appropriate -- if one wants to compare two hypothesis (dark matter vs stochastic spacetime), one needs to consider similar settings. In this case, one should consider the full dynamical cosmological setting on sub-horizon scales, and compare the contribution to the CMB power spectrum due to dark matter, and due to stochastic fluctuations. Or compare dark matter vs stochastic fluctuations in Schwarszchild. Using the power spectrum imprinted into the CMB by dark matter during the evolution of our universe, and comparing that to the power spectrum of stochastic spacetime fluctuations in Schwarszchild, which is a stationary spacetime, is not an appropriate comparison. We initiate work on deriving the power spectrum in cosmology in \cite{UCLcosmo}.

{\bf\noindent Comparison to tabletop experiments:} Next the authors consider the two-point correlation function of the stochastic fluctuations. They are given in~\cite{oppenheim2023gravitationally,grudka2023renormalisable}, and related to that found in the context of quadratic gravity~\cite{stelle1978classical}. In the non-relativistic setting, it is given, up to a constant, by~\cite{oppenheim2023gravitationally,grudka2023renormalisable}
\begin{align*}
	G(x,x'):=& \langle \Phi(x)\Phi(x')\rangle-\langle \Phi(x)\rangle\langle\Phi(x')\rangle
	\nonumber\\
	=&-\frac{G_N^2}{8 \pi(1-\beta) D_0}|\vec{x}-\vec{x}'|\delta(t,t').
\end{align*}
The acceleration covariance matrix can be obtained from this
\begin{align}
	\frac{\partial^2 G(x,x')}{ \partial{x_i}\partial{x'_j}}=\frac{G_N^2}{8\pi(1-\beta) D_{0}} \left( \frac{\delta_{ij} |x - x'|^2 - (x_i - x'_i)(x_j - x'_j)}{|x - x'|^3} \right)\delta(t,t')
\end{align}
and by setting $i=j$ and summing, we obtain the variance of the acceleration $\sigma^2_a(x,x'):=\langle\nabla\Phi(x)\cdot\nabla\Phi(x')\rangle-\langle\nabla\Phi(x)\rangle\cdot\langle\nabla\Phi(x')\rangle$
\begin{align}
	\sigma^2_a(x,x')=\frac{G_N^2}{4\pi(1-\beta) D_{0}}\frac{\delta(t,t')}{|\vec{x}-\vec{x'}|}
\end{align}

The authors of \cite{hertzberg2024comment} then claim that the theory is not well-defined, because the two-point function is singular as $x'\to x$.  In fact, the pure gravity theory is renormalisable~\cite{grudka2023renormalisable}, and the same divergent behaviour is typically seen in the corresponding quantities in quantum field theory. Just as in quantum field theory, the physically relevant quantity is $\sigma^2_a(x,x')$ smeared over a test mass. This is the physically relevant quantity that's measured in a lab. It is also the case that one never makes an instantaneous measurement, but rather, a measurement over some time. One therefore needs to compute, either the time-averaged acceleration over a test mass, or as is more common in precision measurements of this kind, consider the Fourier transform of $a^2(t)$ to find the spectral density.  This quantity has units of $\unit{(m/s^2)^2/Hz}$ -- the variance decreases the longer you average. In this way, one computes the variance per $Hz$ as measured in experiments. Such a calculation was done in \cite{oppenheim2023gravitationally} for some realisations of the theory, and we include a similar updated calculation in Appendix \ref{sec:comparison}. 

Instead, the authors of  \cite{hertzberg2024comment} make a comparison to an absolute measurement of acceleration $2\times10^{-9}\,\unit{m/s^2}$ found in \cite{hoyle2004submillimeter}. But this measurement is averaged over the course of day, and they measure the variance at the resonant frequency. Perhaps more importantly,
while the authors acknowledge that $D_{0,T}=\eps D_0$ for some infinitesimal time $\eps$, this is ignored.  Relating $D_0$ and $D_{0,T}$ cannot be made independent of this choice.  
The anomalous and static contributions ought to be treated separately. For static contributions, it's appropriate to take $\eps$ to be the response time and check if it's consistent with the dynamics as initiated in \cite{UCLcosmo}. One could try to argue that $\eps$ must be taken to be the measurement time used in tabletop experiments, but we are unaware of 
reasons to make this choice.

When this is corrected for, we instead find consistency of our astrophysical results with tabletop experiments, although as we have stressed in v1, one needs to understand the dynamics before one can make strong claims.

On top of this, as we've stressed throughout, comparing the variance in acceleration found in table-top experiments with the variances predicted in the cosmological setting is not straightforward. While it is possible that the variance in acceleration on the Hubble scale may be the same as the variance at millimeter scale, we cannot know without further considerations. As we've noted, the scaling will depend on various factors which we expect future works to clarify, and in particular on the RG flow of the coupling constants or whether fluctuations have mass. The dynamics also plays a role here in how one performs renormalisation. A local fluctuation which is expanded to the Hubble scale during inflation vs the mechanism suggested by Eq. \eqref{eq: NewtVar} where the local variance in acceleration plays the role of the cosmological constant vs a long wavelength fluctuation, would likely all result in different RG flows.

{\bf\noindent Using boundary conditions to remove correlations:} 
In \cite{hertzberg2024comment}, boundary conditions are imposed which fix $\gamma_1$ and $\gamma_2$. It is then claimed that this shows there is no correlation between $\gamma_1$ and $\gamma_2$. It is certainly true that if one fixes random variables, one removes any correlation between them. Two bits with value $00$ are not correlated (learning the value of the first bit tells you nothing new about the second bit), $11$ is also not correlated. But the distribution which is a statistical mixture of $00$ and $11$ does contain correlations between the two bits since learning one tells you what the other one is. If one fixes a random variable, its correlations will be removed, but there is no reason to fix them unless either one has learned the value of one of the random variables or the theory itself has fixed it. In our context, the linear theory does not fix $\gamma_1$ or $\gamma_2$, instead we have a probability distribution over them. If one of them is later determined, either by non-linearities or by the dynamics, then this favours also finding the other contribution.

Again, the example of Brownian motion with a wall, given in Appendix \ref{sec:Brownian}, is a very helpful analogy. If at time $t=0$ we know that the particle is at position $q_0=\alpha_0$, with velocity $\dot{q}_0=\alpha_2$, and if at some later time $t$ (analogous to the ''boundary'' or $r_{max}$ in the gravity case) we find the particle has some $q_f(t)$, then its average acceleration $\ddot{q}$, will be correlated with it's average jerk $\dddot{q}$ (see Eq \eqref{eq:Qmpp})
\begin{equation*}
	q_{MPP}(t)=\alpha_0 + \alpha_1 t+\frac{1}{2}\alpha_2 t^2 + \frac{1}{6}\alpha_3 t^3.
\end{equation*}
and Eq.~\eqref{eq:Qmppaction}). We also don't have to condition on $q_f(t)$ -- a correlation will still exist albeit weaker. In fact, it's initial acceleration and jerk will also be correlated once we condition on $q(t)$, since we are only fixing $\dot{q}_0$ and $\ddot{q}_0$. If on the other hand, at time $t$, we also condition on both the particle's acceleration and jerk, then no correlation exist, but that's always the case with random variables. 

{\bf\noindent Spherically symmetric fluctuations vs fluctuations which are spherically symmetric on average:} The authors also critique the assumption of stochastic fluctuations in time which are spherically symmetric. But this is not something considered by us. Here the anomalous contributions are taken to be stationary, and only spherically symmetric on average.  This is almost certainly true since the matter distribution is spherically symmetric. The average potential and the variances of the distribution will almost certainly be spherically symmetric. How close any individual realisation is to spherical symmetry will depend on the dynamics and how the fluctuations arise. For slow dynamics, one anticipates that the symmetry of the anomalous contributions will be similar to the symmetry of the matter distribution. As with the example of Brownian motion, the  $q_{MPP}(t)$ is a useful tool for understanding the average path (or the most probably path) that the particle takes. But no one would suggest that one is only considering particles with trajectories given by the $q_{MPP}(t)$ with constant $\alpha_i$.

\end{document}